\documentclass[11pt,a4paper]{amsart}
\usepackage{fullpage}
\usepackage{graphicx}
\usepackage{subcaption}
\usepackage{color}
\usepackage{amsmath}
\usepackage{amssymb}
\usepackage{amscd}
\usepackage{bbm}
\usepackage{float}
\usepackage{framed}
\usepackage[ruled,vlined]{algorithm2e}

\usepackage{fancyhdr,a4wide}
\usepackage{amsthm}

\newcommand{\R}{\mathbb{R}}
\newcommand{\inr}[1]{\bigl< #1 \bigr>}

\newcommand{\N}{\mathbb{N}}

\newcommand{\E}{\mathbb{E}}

\newcommand{\eps}{\varepsilon}

\newcommand{\vertiii}[1]{{\left\vert\kern-0.25ex\left\vert\kern-0.25ex\left\vert #1
    \right\vert\kern-0.25ex\right\vert\kern-0.25ex\right\vert}}

\newtheorem{Theorem}{Theorem}[section]
\newtheorem{Lemma}[Theorem]{Lemma}

\newtheorem{Definition}[Theorem]{Definition}

\newtheorem{Remark}[Theorem]{Remark}

\numberwithin{equation}{section}

\def \proof {\noindent {\bf Proof.}\ \ }

\def \endproof
{{\mbox{}\nolinebreak\hfill\rule{2mm}{2mm}\par\medbreak}}

\newcommand{\bP}{\mathbb{P}}

\def\IND{\mathbbm{1}}

\allowdisplaybreaks

\newcommand{\la}{\lambda}
\newcommand{\del}{\delta}
\newcommand{\al}{\alpha}

\newcommand{\ti}{\times}

\newcommand{\thres}{\tau_{\operatorname{thres}}}
\newcommand{\noise}{\nu_{\operatorname{noise}}}
\newcommand{\bnoise}{\bar{\nu}_{\operatorname{noise}}}
\newcommand{\sign}{\operatorname{sign}}
\newcommand{\co}{\operatorname{conv}}

\title{Robust one-bit compressed sensing with partial circulant matrices}
\author{Sjoerd Dirksen}
\thanks{Lehrstuhl C f{\"u}r
Mathematik (Analysis), RWTH Aachen University, dirksen@mathc.rwth-aachen.de}

\author{Shahar Mendelson}

\thanks{Mathematical Sciences Institute, The Australian National University, shahar.mendelson@anu.edu.au}

\begin{document}

\setlength{\parskip}{5pt}

\begin{abstract}
\noindent We present optimal sample complexity estimates for one-bit compressed sensing problems in a realistic scenario: the procedure uses a structured matrix (a randomly sub-sampled circulant matrix) and is robust to analog pre-quantization noise as well as to adversarial bit corruptions in the quantization process. Our results imply that quantization is not a statistically expensive procedure in the presence of nontrivial analog noise: recovery requires the same sample size one would have needed had the measurement matrix been Gaussian and the noisy analog measurements been given as data.
\end{abstract}

\maketitle

\section{Introduction}

The quantization of analog signals to a finite number of bits is an essential step in many signal processing problems: it allows one to digitally transmit, process, and reconstruct signals. In \emph{quantized compressed sensing} the focus is on the recovery of low-complexity signals (e.g., signals that have a sparse representation in a given basis) from their quantized measurements. Such recovery problems are natural, appear frequently in real-world applications, and have been studied extensively in recent years (see, for example, the survey \cite{Dir18}).
\par
A very popular model in quantized compressed sensing is \emph{one-bit compressed sensing}. In this setup, the unknown signal is a (sparse) vector $x \in \R^n$ and linear measurements of the signal are generated using a \emph{measurement matrix} $A\in\R^{m\ti n}$ where $m\ll n$. To make the model realistic, the $m$ linear measurements $(Ax)_i, \ 1 \leq i \leq m$ are corrupted by (random) noise, resulting in the analog measurement vector $Ax+\noise$. Then, each noisy measurement, that is, each coordinate of the vector $Ax+\noise$, is quantized into a single bit by comparing it to a threshold. During this quantization process corruption may occur again, leading to several `sign flips'. In other words, if we set $\thres\in \R^m$ to be the vector whose coordinates are the quantization thresholds, $\text{sign}(\cdot)$ is the sign function applied element-wise, and
\begin{equation} \label{eqn:1-bitModel}
q=\text{sign}(Ax + \noise + \thres),
\end{equation}
then the data one actually receives is a corrupted vector $q_{\operatorname{corr}} \in \{-1,1\}^m$, obtained from $q$ by several (possibly adversarial) sign changes.

In realistic situations, one has no control on the noise vector $\noise$ which determines the pre-quantization (analog) noise, nor on the sign changes that may occur during quantization. The one component that can be controlled is the vector $\thres$ which determines the thresholds used in the quantization process. As it happens, if the quantization thresholds are either fixed or are random and independent, then the one-bit quantizer $\text{sign}(\cdot + \thres)$ can be implemented very efficiently; it should therefore come as no surprise that it is popular in engineering literature (see e.g.\ \cite{BoB08,MoH15}).

\par

Our main interest here is to explore one-bit compressed sensing in realistic problems and to present realistic solutions for such problems. This requires addressing two core issues:

\par

\noindent{\bf \underline{Noise.}} It is a fact of life that noise plays a significant role in real-world problems. Indeed, one encounters noise at the analog, pre-quantization phase and also during the quantization process. What plays a crucial role is the noise level one faces: when the analog noise vector $\noise$ has iid coordinates, the noise level is captured by the variance of the coordinates; and during quantization that noise level is the maximal number of bits that can be `flipped'.

In realistic problems the two noise levels can be substantial: the variance of the coordinates of $\noise$ can be some constant that has nothing to do with the required reconstruction accuracy, and at the same time, the number of sign changes that may occur during quantization can be a fixed proportion of $m$. As a result, solutions to realistic recovery problems must be based on procedures that are robust to the effect of significant noise levels.

\par

\noindent{{\bf \underline {Structured measurement matrices.}}  In classical (`unquantized') compressed sensing, it is well known that optimal reconstruction guarantees are enjoyed by completely random measurement matrices, such as the standard Gaussian matrix. Unfortunately, such matrices are extremely difficult to `realize' in practice, as real-world measurement schemes are subject to physical constraints, and those constraints lead to highly structured measurement matrices. Thus, if one is looking for a realistic procedure, \emph{one must use structured measurement matrices}.

\vskip0.4cm

Despite the popularity of one-bit compressed sensing, the current state-of-the-art falls well-short of addressing realistic scenarios. Firstly, all existing results deal with problems that are either noiseless or have an analog noise level (i.e. the variance of the coordinates of $\noise$) that is small relative to the wanted reconstruction accuracy, making the problem de-facto noiseless (see \cite{Dir18} for an overview of these results). Moreover, the issue of post-quantization bit corruptions is typically not dealt with at all (two exceptions are \cite{DiM18a,PlV13}).

Secondly, almost all the relevant work has focused on a standard Gaussian measurement matrix. The reason is that one-bit compressed sensing can very easily \emph{fail} when using a non-Gaussian matrix---even if that matrix is known to perform optimally in classical compressed sensing. Indeed, when all the thresholds are set to $0$ (the scenario studied, e.g., in \cite{ALP14,DJR17}) there are $2$-sparse vectors that are `far away' from one another and still cannot be distinguished based on their quantized Bernoulli measurements (see \cite{ALP14}). Recently it was shown in \cite{DiM18a} that one-bit compressed sensing is possible for a large class of non-Gaussian measurement matrices---though still with iid rows---by invoking \emph{dithering}; that is, by selecting well-designed random thresholds for the quantization process. Unfortunately, while \cite{DiM18a} extends the scope of the method beyond the Gaussian case, it still does not address the key difficulty: that measurement matrices with iid rows are rather useless when it comes to the study of realistic problems.

Intuitively, the constraint that the measurement matrix should be structured is a major obstacle, because the behaviour of a structured measurement matrix is likely to be less favourable than of a fully random one. Thankfully, not all is lost: there are examples in classical compressed sensing literature which show that near-optimal sample complexities can still be achieved using \emph{structured random matrices}; that is, using matrices that are generated by injecting some (minimal) randomness into realistic measurement models.

A very popular family of structured random matrices are the randomly sub-sampled circulant matrices, where the resulting measurements amount to randomly sub-sampling the discrete circular convolution of the unknown signal with a random pulse (for more details see below). This method of measurement is very popular and is used extensively in applications, ranging from SAR radar imaging through optical imaging and channel estimation (see e.g.\ \cite{Rom09} and the references therein).

\par
The goal of this article is to resolve the two issues that are at the heart of real-world problems: that the measurement matrix must be structured and that the given measurements are noisy. Indeed,

\begin{framed}
We establish an optimal (up to logarithmic factors) one-bit sparse recovery procedure for realistic problems: the pre-quantization noise can be high; during the quantization process a large fraction of the signs may change in an adversarial way; and the measurement matrix is structured---a \emph{randomly sub-sampled circulant matrix}.
\end{framed}
\par

Before we formulate our main results, let us introduce some notation, beginning with the measurement matrix we use. Let $\xi\in \R^n$ be a random vector with independent, mean-zero, unit variance, $L$-subgaussian\footnote{Recall that a centred random variable is $L$-subgaussian if for every $p \geq 2$, $\|\xi\|_{L_p} \leq L\sqrt{p} \|\xi\|_{L_2}$.} coordinates. Let $\Gamma_{\xi}$ be the circulant matrix generated by $\xi$; that is, the $j$-th row of $\Gamma_\xi$ is $(\xi_{j \ominus k})_{k=1}^n$ where $\ominus$ is subtraction mod $n$. Consider independent $\{0,1\}$-valued random variables $\del_1,\ldots \del_n$ with mean $\delta=m/n$, which are independent of $\xi$; let $I=\{i\in [n] \ : \ \delta_i=1\}$ and set $R_I$ to be the associated restriction operator. The measurement matrix we use is $A =R_I\Gamma_{\xi}$, i.e., a randomly sub-sampled circulant matrix whose rows are chosen from the rows of $\Gamma_\xi$ according to the selectors $(\delta_i)_{i=1}^n$.

\par

Next, let us turn to the analog noise vector. Let $\nu_1,...,\nu_m$ be independent copies of a random variable $\nu$ (that need not be centred) which are also independent of $(\delta_i)_{i=1}^n$ and $\xi$. Thus, the noise vector $\noise=(\nu_i)_{i=1}^m$ consists of iid coordinates, but can have a nontrivial `drift'.

The choice of the thresholds used in the quantization process turns out to be of central importance. The thresholds are defined using $\tau_1,...,\tau_m$, which are independent copies of a centred random variable $\tau$. Set $\thres=(\tau_i)_{i=1}^m$ and assume that $\thres$ is independent of $(\delta_i)_{i=1}^n$, $\xi$, and $\noise$.

Finally, we assume that at most $\beta m $ bits are corrupted arbitrarily during quantization for some parameter $0<\beta<1$. Thus, if $q$ is as in \eqref{eqn:1-bitModel} (i.e., $q$ is the `perfect' quantization of the vector of noisy analog measurements) and $d_H$ denotes the Hamming distance, then instead of $q$ one observes a corrupted measurement vector $q_{\operatorname{corr}} \in \{-1,1\}^m$ which satisfies $d_H(q_{\operatorname{corr}},q)\leq \beta m$.

\par

Throughout we assume that the unknown signal is $s$-sparse and denote by $\Sigma_{s,n}$ the set of $s$-sparse vectors in the Euclidean unit ball in $\R^n$. The recovery procedure we use is
\begin{equation} \label{eqn:progIsomorphicIntro}
\max_{z\in \Sigma_{s,n}} \frac{1}{m}\inr{q_{\operatorname{corr}},Az} - \frac{1}{2\lambda} \frac{\|\Gamma_{\xi} z\|_2^2}{n},
\end{equation}
and its performance is described in the following theorem, which is the main result of this article.
\begin{Theorem} \label{thm:isomorphic}
For $L \geq 1$ there exist constants $c_1,...,c_4$ that depend only on the subgaussian constant $L$, and poly-logarithmic factors $\gamma_1,\gamma_2$ satisfying
$$
\gamma_1\leq \log(s)\log(n), \qquad \gamma_2\leq \log(n)\log\log(n)
$$
such that the following holds. Fix $0<\rho<1$ and assume that $\nu$ is $L$-subgaussian and that $|\E \nu| \leq c_1 \rho$. Set $\bar{\nu}=\nu - \E \nu$, let
$$
\lambda \geq c_2 \gamma_1 \max\{\|\bar{\nu}\|_{L_2},1\} \log(e\gamma_1^2 \max\{\|\bar{\nu} \|_{L_2},1\}/\rho),
$$
and set $\beta$ such that
$$
\beta \sqrt{\log(e/\beta)} \leq \frac{c_3}{\gamma_1 \gamma_2} \cdot \frac{\rho}{\lambda}.
$$
Let $\tau$ be uniformly distributed on $[-\lambda,\lambda]$ and set
$$
m \geq c_4 \gamma_1^2 \gamma_2^2 \frac{\lambda^2 s \log(en/s)}{\rho^2}.
$$
Then, with probability at least $1-(\frac{s}{n})^2$, for any $s$-sparse $x\in \R^n$ with $\|x\|_2\leq 1$, any solution $x^{\#}$ to \eqref{eqn:progIsomorphicIntro} satisfies $\|x^{\#}-x\|_2\leq \rho$.
\end{Theorem}

\begin{Remark}
As the proof of Theorem~\ref{thm:isomorphic} shows, the probability estimate can be improved to $1-(s/n)^\zeta$ for any $\zeta \geq 2$ at a price of modified constants $c_1,...,c_4$.
\end{Remark}

The number of bits that can be safely corrupted during quantization without damaging the accuracy is, up to logarithmic terms, the best that one can hope for in the setting of Theorem~\ref{thm:isomorphic} --- it is possible to show that if one aims for recovery with accuracy $\rho$, then no more than $\sim \rho m$ of the bits can be corrupted in an adversarial way during quantization (up to logarithmic terms). But what is more striking is that Theorem~\ref{thm:isomorphic} is (almost) optimal in a rather strong (minimax) sense, as the next result shows.
\begin{Theorem} \label{thm:lower}
Let $\nu$ be a centred Gaussian random variable, set $A$ to be a (random) measurement matrix that satisfies, with probability at least $0.95$,
\begin{equation} \label{eq:cond-on-A-lower-bound}
\|Ax\|_2 \leq \kappa \sqrt{m} \|x\|_2, \qquad \text{for all } x \in \Sigma_{s,n}.
\end{equation}
Let $\Psi$ be any recovery procedure such that, for every fixed $x \in \Sigma_{s,n}$, when receiving as data the measurement matrix $A$ and the noisy linear measurements $((Ax)_i+\nu_i)_{i=1}^m$, $\Psi$ returns $x^\sharp$ that satisfies $\|x^\sharp-x\|_2 \leq \rho$ with probability $0.9$. Then
$$
m \geq c\kappa^{-2} \|\bar{\nu}\|_{L_2}^2 \frac{s \log(en/s)}{\rho^2}.
$$
\end{Theorem}

The meaning of Theorem~\ref{thm:lower} is that even if one receives the noisy analog linear measurements prior to quantization, and is then free to use those measurements as one sees fit, the sample size required for recovery with accuracy $\rho$ is at least $\|\bar{\nu}\|_{L_2}^2 s\log(en/s)/\rho^2$. In light of Theorem \ref{thm:isomorphic}, and perhaps contrary to intuition, this means that quantization is not a statistically expensive procedure in the presence of nontrivial analog noise: by using one-bit quantization with uniformly distributed thresholds, combined with the efficient recovery scheme \eqref{eqn:progIsomorphicIntro}, the recovery performance is the best that one can hope for (up to a poly-logarithmic factor), even if one had been given the complete noisy analog measurements. In particular, sophisticated quantization schemes that collect more bits per measurement (see e.g., \cite{DJR17,XuJ18}) and/or quantize in an adaptive way (e.g., the methods in \cite{FKS17,HuS18}) are not effective in realistic problems in which the analog noise level is nontrivial.

The situation in the less realistic scenario of a low analog noise level is entirely different and an appropriate version of Theorem~\ref{thm:isomorphic} may be used to achieve the optimal sample complexity in that scenario as well (see Section \ref{sec:extensions}).

\begin{Remark}
It is well known that \eqref{eq:cond-on-A-lower-bound} holds with probability $0.95$ for many random measurement matrices studied in compressed sensing if $m\geq c\gamma s \log(en/s)$ and $\gamma$ is a poly-logarithmic factor; in particular, \eqref{eq:cond-on-A-lower-bound} is satisfied when $A$ has iid subgaussian rows or when $A$ is a partial circulant matrix generated by an $L$-subgaussian random vector.
\end{Remark}

\par

The article is organized as follows. In Section~\ref{sec:analysisGen} we analyze the recovery procedure \eqref{eqn:progIsomorphicIntro} for a general matrix $\Gamma$ (and not only for a circulant matrix $\Gamma_{\xi}$) and deduce sufficient conditions on $\Gamma$ that ensure that the procedure is successful. In Section~\ref{sec:circulnat} we verify that the required conditions are satisfied by a subgaussian circulant matrix, thereby completing the proof of Theorem~\ref{thm:isomorphic}. Section~\ref{sec:lower} is devoted to the proof of Theorem~\ref{thm:lower}, and in Section~\ref{sec:extensions} we sketch several extensions of Theorem~\ref{thm:isomorphic}, including its implications for the low noise regime.

\subsection{Notation}

For $k\in \N$ let $[k]=\{1,\ldots,k\}$. $|S|$ denotes the cardinality of a set $S$. Given $x \in \R^n$, set $\|x\|_0=|\{i \in [n] \ : \ x_i\neq 0\}|$; let  $\Sigma_{s,n}=\{x\in \R^n \ : \ \|x\|_0\leq s, \ \|x\|_2\leq 1\}$ be the set of $s$-sparse vectors in the Euclidean unit ball; $\|x\|_p$ denotes the $\ell_p$-norm and put $B_p^n = \{x \in \R^n : \|x\|_p \leq 1\}$.

Recall that $d_H$ is the (unnormalized) Hamming distance on the discrete cube and for a centred random variable $\xi$ set
$$
\|\xi\|_{\psi_2} = \sup_{p\geq 1}\frac{\|\xi\|_{L_p}}{\sqrt{p}}.
$$
Finally, $c$ and $C$ denote absolute constants; their value may change from line to line. $c_\alpha$ or $C(\alpha)$ denotes a constant that depends only on the parameter $\alpha$. We write $a\lesssim_{\al} b$ if $a\leq C_{\al} b$, and $a\simeq_{\al} b$ means that both $a\lesssim_{\al} b$ and $a\gtrsim_{\al} b$ hold.

\section{{Analysis of the recovery method}} \label{sec:analysisGen}

In what follows $\Gamma$ is an ${n\times n}$ matrix, and the measurement matrix we consider is obtained by randomly selecting rows of $\Gamma$ using independent $\{0,1\}$-valued random variables (selectors) $\del_1,\ldots,\del_n$ with mean $\delta=m/n$. Hence, $A$ is defined by
$$
A z = \sum_{i=1}^n \delta_i \inr{\Gamma z,e_i} e_i.
$$
Observe that the number of measurements may be slightly different from $m$. It is the cardinality of the set $\{i \in [n]: \delta_i=1\}$, which, by the Chernoff bound, concentrates in $[m/2,3m/2]$ with probability at least $1-e^{-cm}$ .

\par
Our proof of Theorem~\ref{thm:isomorphic} consists of two independent components. We first show that the program \eqref{eqn:progIsomorphicIntro} succeeds if $\Gamma$ behaves `as if it were a Gaussian matrix' in two distinct ways:
\begin{itemize}
\item It acts as an isomorphism on sparse vectors, i.e., for suitable constants $0<c<C<\infty$,
\begin{equation}
\label{eqn:isomorphicIntro}
c\|x\|_2\leq \frac{1}{\sqrt{n}}\|\Gamma x\|_2\leq C\|x\|_2, \qquad \text{for all} \ x\in \Sigma_{s,n}.
\end{equation}
\item Any vector in $\Gamma(\Sigma_{s,n})$ satisfies a \emph{growth property}: that is, for every $x\in \Sigma_{s,n}$,
\begin{equation} \label{eq:growth-0}
\|\Gamma x\|_{[k]} \leq \gamma_1  \sqrt{\frac{k\log(en/k)}{n}} \|\Gamma x\|_{2}, \qquad \text{for all} \ k\geq s
\end{equation}
where $\gamma_1$ is a poly-logarithmic factor in $s$ and $n$. Here, for a vector $w\in \R^n$, $w^*$ is the non-increasing rearrangement of $(|w_i|)_{i=1}^n$ and
$$
\|w\|_{[k]} = \Bigl(\sum_{i=1}^k (w_i^*)^2\Bigr)^{1/2}
$$
is the $\ell_2$-norm of the $k$-largest coordinates.
\end{itemize}

\par

In the second part of the proof we show that a random circulant matrix generated by a subgaussian random vector exhibits the Gaussian-like behaviour \eqref{eqn:isomorphicIntro} and \eqref{eq:growth-0} with high probability, despite the rather `limited randomness' such a matrix has. This surprising feature is discussed in detail in Section~\ref{sec:circulnat}.

To start our analysis fix the matrices $\Gamma$ and $A$; the given set $T\subset \R^n$; and the corrupted vector of quantized measurements $q_{\operatorname{corr}}$. Define the functional $\phi:\R^n\to \R$ by
\begin{equation} \label{eqn:FunIsomorphic}
\phi(z)=\frac{1}{m}\inr{q_{\operatorname{corr}},Az} - \frac{1}{2\lambda} \frac{\|\Gamma z\|_2^2}{n}.
\end{equation}
The recovery procedure we explore is
\begin{equation}
\label{eqn:progIsomorphic}
\max_{z \in T} \phi(z).
\end{equation}
Although our focus is on the set $T=\Sigma_{s,n}$ (leading to the program \eqref{eqn:progIsomorphicIntro}), the method of analysis presented here can be used to study \eqref{eqn:progIsomorphic} for other sets $T$, most notably $T=\sqrt{s} B_1^n\cap B_2^n$. The latter set is used in approximate sparse recovery problems (see more details in Section~\ref{sec:extensions}).

\par

To establish Theorem~\ref{thm:isomorphic} consider the `excess functional' $\phi(z)-\phi(x)$. In what follows we show that for the wanted reconstruction error $\rho$, and using $m$ measurements, one can ensure that $\phi(z)-\phi(x) < 0$ whenever $x,z \in T$ and $\|x-z\|_2 \geq c\rho$.  That implies that, for any $x\in T$, any solution $x^{\#}$ to \eqref{eqn:progIsomorphic} satisfies $\|x^{\#}-x\|_2 \leq c\rho$.

\subsection{Decomposition of the excess risk}

The first step in the proof is a decomposition of the excess functional. Observe that
 \begin{align}
\label{eqn:mainDecompExcess}
\phi(z)-\phi(x) & = \frac{1}{m} \inr{q_{\operatorname{corr}},Az-Ax} -\frac{1}{2\lambda}\frac{\|\Gamma z\|_2^2}{n} + \frac{1}{2\lambda}\frac{\|\Gamma x\|_2^2}{n} \nonumber
\\
& = \frac{1}{m} \inr{q_{\operatorname{corr}}-\sign{(Ax+\noise+\thres)},A(z-x)} \nonumber
\\
& + \frac{1}{m} \left(\inr{\sign{(Ax+\noise+\thres)},A(z-x)} \right. \nonumber \\
& \qquad \qquad \qquad \qquad \qquad - \left. \E_{\delta \otimes \nu \otimes \tau} \inr{\sign{(Ax+\noise+\thres)},A(z-x)} \right) \nonumber
\\
& + \frac{1}{m}\E_{\delta \otimes \nu \otimes \tau} \inr{\sign{(Ax+\noise+\thres)},A(z-x)} -\frac{1}{2\lambda}\frac{\|\Gamma z\|_2^2}{n} + \frac{1}{2\lambda}\frac{\|\Gamma x\|_2^2}{n} \nonumber
\\
& =: (1)+(2)+(3),
\end{align}
where $\E_{\delta \otimes \nu \otimes \tau}$ is the expectation with respect to $(\delta_i)_{i=1}^n$, $(\nu_i)_{i=1}^n$ and $(\tau_i)_{i=1}^n$, $\noise=(\delta_i \nu_i)_{i=1}^n$ and $\thres=(\delta_i \tau_i)_{i=1}^n$.

The goal is to use this decomposition and find a constant $C$ and a high probability event on which, for every $x \in T$ and $z \in T$ that satisfy $\|x-z\|_2 \gtrsim \rho$,
\begin{equation} \label{eq:path}
|(1)|  \leq C\|x-z\|_2^2; \ \ |(2)| \leq C\|x-z\|_2^2 \ \ {\rm and} \ \ (3) \leq -4C\|x-z\|_2^2,
\end{equation}
implying that $\phi(z)-\phi(x) \leq -2C\|x-z\|_2^2$ when $\|x-z\|_2 \gtrsim \rho$.

\vskip0.4cm

Writing $q_{\operatorname{corr}} =(q_i)_{i=1}^n$, the three terms in \eqref{eqn:mainDecompExcess} are

\begin{equation} \label{eq:(1)-using-Gamma}
(1) = \frac{1}{m} \sum_{i=1}^n \delta_i \bigl(q_i -\sign((\Gamma x)_i+\nu_i +\tau_i)\bigr) \cdot \bigl(\Gamma(z-x)\bigr)_i;
\end{equation}
\begin{align} \label{eq:(2)-using-Gamma}
(2) = \frac{1}{m}\sum_{i=1}^n \Bigl( & \delta_i \sign((\Gamma x)_i+\nu_i +\tau_i) \cdot \bigl(\Gamma(z-x)\bigr)_i
\\
& -  \E_{\delta \otimes \nu \otimes \tau} \ \delta_i \sign((\Gamma x)_i+\nu_i +\tau_i) \cdot \bigl(\Gamma(z-x)\bigr)_i\Bigr); \nonumber
\end{align}
and
\begin{equation} \label{eq:(3)-using-Gamma}
(3) = \frac{1}{m}\sum_{i=1}^n \E_{\delta \otimes \nu \otimes \tau} \ \delta_i \sign((\Gamma x)_i+\nu_i +\tau_i) \cdot \bigl(\Gamma(z-x)\bigr)_i -\frac{1}{2\lambda}\frac{\|\Gamma z\|_2^2}{n} + \frac{1}{2\lambda}\frac{\|\Gamma x\|_2^2}{n}.
\end{equation}

\vskip0.4cm

\noindent {\bf{\underline{The term \eqref{eq:(3)-using-Gamma}}}}

\vskip0.4cm

To estimate \eqref{eq:(3)-using-Gamma} it suffices to show that for every $x \in T$ and every $z \in T$ that satisfies $\|z-x\|_2 \gtrsim \rho$,
\begin{framed}
\begin{equation} \label{eq:est-on-(3)-1}
\frac{1}{m}\sum_{i=1}^n \E_{\delta \otimes \nu \otimes \tau} \delta_i \sign((\Gamma x)_i+\nu_i +\tau_i) \bigl(\Gamma(z-x)\bigr)_i \leq \frac{1}{\lambda} \frac{\inr{\Gamma x,\Gamma (z-x)}}{n} + \frac{\rho}{16 \lambda} \frac{\|\Gamma (z-x)\|_2}{\sqrt{n}}.
\end{equation}
\end{framed}

Indeed, if that is the case then
\begin{align*}
(3) \leq  & \frac{1}{\lambda n} \left(\inr{\Gamma x,\Gamma (z-x)} - \frac{\|\Gamma z\|_2^2}{2} + \frac{\|\Gamma x\|_2^2}{2} \right) + \frac{\rho }{16 \lambda}\frac{\|\Gamma(z-x)\|_2}{\sqrt{n}}
\\
= & \frac{1}{2\lambda} \frac{\|\Gamma(z-x)\|_2}{\sqrt{n}} \left(- \frac{\|\Gamma (z-x)\|_2}{\sqrt{n}}+\frac{\rho}{8} \right)=(*).
\end{align*}
Hence, if the matrix $\Gamma$ satisfies a \emph{small-ball property}, namely, that there is a constant $0<\kappa<1$ such that for every $x,z \in T$,
\begin{equation}
\label{eqn:SBGammaAss}
\frac{\|\Gamma(z-x)\|_2}{\sqrt{n}} \geq \kappa \|z-x\|_2,
\end{equation}
and if $\|x-z\|_2 \geq \rho/4\kappa$, then
\begin{equation} \label{eq:est-on-(3)-2}
(*) \leq -\frac{1}{2\lambda} \frac{\|\Gamma(z-x)\|}{\sqrt{n}} \cdot \frac{\kappa}{2} \|z-x\|_2 \leq -\frac{\kappa^2}{4\lambda} \|z-x\|_2^2,
\end{equation}
which is the wanted estimate. Of course, it suffices if \eqref{eqn:SBGammaAss} holds only when $\|x-z\|_2 \gtrsim \rho$.

\vskip0.4cm
\noindent {\bf{\underline{The term \eqref{eq:(2)-using-Gamma}}}}
\vskip0.4cm
If we set
$$
\Psi(x,y)= \frac{1}{m}\sum_{i=1}^n \delta_i \sign((\Gamma x)_i+\nu_i +\tau_i) \cdot \bigl(\Gamma(y/\|y\|_2^2) \bigr)_i,
$$
then
$$(2)=\|z-x\|_2^2 \ \bigl(\Psi(x,z-x)-\E_{\delta \otimes \nu \otimes \tau}\Psi(x,z-x)\bigr).$$
Observe that
\begin{align*}
& \sup_{x \in T} \ \sup_{\{z \in T, \ \|x-z\|_2 \geq \rho\}} \bigl|\Psi(x,z-x)-\E_{\delta \otimes \nu \otimes \tau}\Psi(x,z-x)\bigr|
\\
& \qquad \leq \sup_{x \in T} \ \sup_{\{y \in T-T, \ \|y\|_2 \geq \rho\}} \bigl|\Psi(x,y)-\E_{\delta \otimes \nu \otimes \tau}\Psi(x,y)\bigr|
\end{align*}
and the wanted estimate on \eqref{eq:(2)-using-Gamma} follows once one identifies a high probability event on which, for every $x \in T$ and any $y \in T-T$ such that $\|y\|_2 \geq \rho$,
$$
\bigl|\Psi(x,y)-\E_{\delta \otimes \nu \otimes \tau}\Psi(x,y)\bigr| \leq \frac{1}{16 \lambda}.
$$
Such an estimate calls for a `{\emph star-shape argument}': if $f:\R^n\to \R_+$ is positive homogeneous and $W\subset \R^n$ is star-shaped around $0$, i.e., $\theta w\in W$ for all $w\in W$ and $0<\theta<1$, then
$$
\sup_{\{w\in W \ : \ \|w\|_2\geq \rho\}} f(w/\|w\|_2^2) \leq \sup_{\{w\in W \ : \ \|w\|_2=\rho\}} f(w)/\rho^2.
$$

Observe that for every fixed $x$, $(\delta_i)_{i=1}^n$, $(\nu_i)_{i=1}^n$, and $(\tau_i)_{i=1}^n$ the function
$$
f(w)=\Bigl|\frac{1}{m}\sum_{i=1}^n \bigl(\delta_i \sign((\Gamma x)_i+\nu_i +\tau_i) \cdot (\Gamma w)_i - \E_{\delta \otimes \nu \otimes \tau} \delta_i \sign((\Gamma x)_i+\nu_i +\tau_i) \cdot (\Gamma w)_i \bigr)\Bigr|
$$
is positive homogenous in $w$, and by the star-shape argument
$$
\sup_{\{y \in T-T, \ \|y\|_2 \geq \rho\} } f(y/\|y\|_2^2) \leq \sup_{\{y \in {\rm star}(T-T), \ \|y\|_2 = \rho\}} \frac{f(y)}{\rho^2},
$$
where for a set $W$ we denote by ${\rm star}(W)$ the set $\{\theta w : 0 \leq \theta \leq 1, \ w \in W\}$.

Therefore, one has to show that with high probability,
\begin{align*}
\sup_{x \in T} \ \sup_{\{y \in {\rm star}(T-T), \|y\|_2 = \rho\} } \Bigl| & \frac{1}{m}\sum_{i=1}^n \Bigl(\delta_i \sign((\Gamma x)_i+\nu_i +\tau_i) \cdot (\Gamma y)_i
\\
& \qquad -\E_{\delta \otimes \nu \otimes \tau} \delta_i \sign((\Gamma x)_i+\nu_i +\tau_i) \cdot (\Gamma y)_i \Bigr)\Bigr| \leq \frac{\rho^2}{16 \lambda}, \nonumber
\end{align*}
which follows by a standard symmetrization argument \cite{GiZ84} once
\begin{framed}
\begin{equation} \label{eq:est-on-(2)-1}
\sup_{x \in T} \ \sup_{\{ y \in {\rm star}(T-T), \ \|y\|_2 = \rho\} }  \Bigl| \frac{1}{m}\sum_{i=1}^n \delta_i \eps_i \sign((\Gamma x)_i+\nu_i +\tau_i) \cdot (\Gamma y)_i \Bigr| \leq \frac{\rho^2}{32 \lambda};
\end{equation}
\end{framed}
\noindent here and throughout, $(\eps_i)_{i=1}^n$ are independent, symmetric $\{-1,1\}$-valued random variables that are independent of $(\delta_i)_{i=1}^n$,  $(\nu_i)_{i=1}^n$, and $(\tau_i)_{i=1}^n$.

\vskip0.4cm
\noindent {\bf {\underline{The term \eqref{eq:(1)-using-Gamma}}}}
\vskip0.4cm
Using an almost identical argument, it suffices to show that with high probability,
\begin{equation} \label{eq:est-on-(1)-1}
\sup_{x \in T} \sup_{\{y \in {\rm star}(T-T), \ \|y\|_2 = \rho\}} \Bigl| \frac{1}{m} \sum_{i=1}^n \delta_i \Bigl(q_i -\sign\bigl((\Gamma x)_i+\nu_i +\tau_i\bigr)\Bigr) \cdot \bigl(\Gamma y\bigr)_i \Bigr| \leq \frac{\rho^2}{16 \lambda}.
\end{equation}
Set $J=\{j : \delta_j =1\}$ and recall that for every target vector $x$ and any realization of $(\delta_i)_{i=1}^n$, $(\nu_i)_{i=1}^n$ and $(\tau_i)_{i=1}^n$ one has that $|\{j \in J : q_j \not = \sign((\Gamma x)_j+\nu_j +\tau_j)\}| \leq \beta m$. Therefore,
$$
\Bigl| \frac{1}{m} \sum_{i=1}^n \delta_i \Bigl(q_i -\sign\bigl((\Gamma x)_i+\nu_i +\tau_i\bigr)\Bigr) \bigl(\Gamma y\bigr)_i \Bigr| \leq \max_{|I| \leq \beta m} \frac{1}{m} \sum_{i\in I} \delta_i |(\Gamma y)_i|
$$
and one has to show that on a high probability event,
\begin{framed}
\begin{equation} \label{eq:est-on-(1)-2}
\sup_{\{y \in {\rm star}(T-T), \ \|y\|_2 = \rho\}} \max_{|I| \leq \beta m} \frac{1}{m} \sum_{i=1}^n \delta_i |(\Gamma y)_i| \leq \frac{\rho^2}{16 \lambda}.
\end{equation}
\end{framed}

\subsection{Controlling the three terms}

Before continuing with the study of the excess loss functional, let us explore the sets $T$ and ${\rm star}(T-T) \cap \rho S^{n-1}$ in the case that we are interested in.
Observe that if $T=\Sigma_{s,n}$, then
\begin{equation}
\label{eqn:starHullSp}
{\rm star}(T-T) \cap \rho S^{n-1} \subset 2\rho \Sigma_{2s,n}.
\end{equation}
Motivated by \eqref{eqn:starHullSp}, that means exploring \eqref{eq:est-on-(3)-1}, \eqref{eq:est-on-(2)-1} and \eqref{eq:est-on-(1)-2} for the pair of sets
\begin{equation} \label{eq:sets-0}
\Sigma_{s,n}, \ \ \ \ \rho \Sigma_{2s,n}.
\end{equation}
As will become clear, the geometry of the images of the two sets under $\Gamma$ is of the utmost importance; specifically, the elements of the images need to satisfy the following fundamental property.

\vskip0.4cm
Given a vector $(x_i)_{i=1}^n$, recall that $(x_i^*)_{i=1}^n$ is the nonincreasing rearrangement of $(|x_i|)_{i=1}^n$ and that for $1 \leq k \leq n$,
$$
\|x\|_{[k]}= \Bigl(\sum_{i \leq k} (x_i^*)^2 \Bigr)^{1/2}.
$$

\begin{Definition} \label{def:growth-property}
A vector $x \in \R^n$ satisfies the growth property with parameters $r$ and $\gamma_1\geq 1$ if for every $r \leq k \leq n$,
\begin{equation} \label{eq:growth}
\|x\|_{[k]} \leq \gamma_1  \sqrt{\frac{k\log(en/k)}{n}} \|x\|_{2}.
\end{equation}
\end{Definition}

\vskip0.4cm

The motivation behind Definition \ref{def:growth-property} is regularity, as vectors that satisfy \eqref{eq:growth} are `well-spread'. Indeed, the contribution  to $\|x\|_2$ by the $k$ largest coordinates of $(|x_i|)_{i=1}^n$ is rather limited unless $k$ is close to $n$. Moreover, while there is little information on how the coordinates $(x_1^*,...,x_r^*)$ are distributed, beyond that the coordinates of $x$ are almost constant and contribute a proportion of $\|x\|_2$. To see that note that if
$$
\gamma_1 \sqrt{\frac{r \log(en/r)}{n}} \leq \frac{1}{2}
$$
then
$$
\Bigl(\sum_{i=r+1}^n (x_i^*)^2\Bigr)^{1/2} \geq \frac{\|x\|_2}{2},
$$
and at the same time, for any $k \geq r$,
\begin{equation} \label{eq:monotone-single}
x_k^* \leq \frac{\|x\|_{[k]}}{\sqrt{k}} \leq \gamma_1 \sqrt{\frac{\log(en/k)}{n}} \|x\|_2.
\end{equation}

\subsection{Proof of \eqref{eq:est-on-(3)-1}}
Recall that by our assumptions, $\bar{\nu}=\nu-\E \nu$ is an $L$-subgaussian random variable and that $\tau$ is distributed uniformly in $[-\lambda,\lambda]$.
\vskip0.3cm
\begin{Theorem} \label{thm:est-on-(3)-sparse}
There exist constants $c_1,c_2,c_3$ and $c_4$ that depend only on $L$ for which the following holds.
Assume that $0<\rho<1$; that
\begin{itemize}
\item[$(a)$] for every $t \in \Sigma_{2s,n}$, $\|\Gamma t\|_2\leq c_1\sqrt{n}$ and $\Gamma t$ satisfies the growth property \eqref{eq:growth} with constants $r=2s$ and $\gamma_1\geq 1$;
\item[$(b)$] also, $| \E \nu | \leq c_2 \rho$,
$$
\rho \geq c_3 \gamma_1^2 \max\{\|\bar{\nu}\|_{L_2},1\}\frac{s}{n}\log(en/s),
$$
and
$$
\lambda \geq c_4 \gamma_1 \max\{\|\bar{\nu}\|_{L_2},1\} \log(e\gamma_1^2 \max\{\|\bar{\nu}\|_{L_2},1\}/\rho).
$$
\end{itemize}
Then for every $x,z \in \Sigma_{s,n}$,
$$
\left|\E_{\delta \otimes \nu \otimes \tau} \frac{1}{m}\sum_{i=1}^n \delta_i  \sign \bigl( (\Gamma x)_i + \nu_i +\tau_i \bigr)\cdot (\Gamma (x-z))_i - \frac{1}{\lambda} \frac{\inr{\Gamma x,\Gamma (z-x)}}{n}\right| \leq \frac{\rho}{16 \lambda} \frac{\|\Gamma (z-x)\|_2}{\sqrt{n}}.
$$
In particular, if for every $t \in \Sigma_{2s,n}$, $\|\Gamma t\|_2/\sqrt{n} \geq \kappa \|t\|_2$, then for any $x,z \in \Sigma_{s,n}$ that satisfy $\|x-z\|_2 \geq 4\rho/\kappa$ one has
$$
(3) \leq -\frac{\kappa^2}{4\lambda} \|z-x\|_2^2.
$$
\end{Theorem}

\vskip0.4cm

The key estimate in the proof of Theorem \ref{thm:est-on-(3)-sparse} is as follows. For every ${w,v} \in \R^n$ set
$$
Z_{w,v} = \frac{1}{m} \sum_{i=1}^n \delta_i v_i \sign ({w_i}+\nu_i + \tau_i)
$$
where $(\delta_i)_{i=1}^n$ are, as always, independent, $\{0,1\}$-valued random variables with mean $\delta =m/n$. Theorem~\ref{thm:est-on-(3)-sparse} is an immediate application of \eqref{eq:est-on-(3)-2} and the following fact, with the choices $w=\Gamma x$ and $v=\Gamma(z-x)$.
\begin{Theorem} \label{thm:Z-wv-est}
There exist constants $c_1$ and $c_2$ that depend only on $L$ for which the following holds. Let $w,v \in \R^n$ satisfy the growth property \eqref{eq:growth} with parameters $r$ and $\gamma_1$. Set $0<\rho<1$ and $0<\theta<1$ such that
\begin{equation} \label{eq:cond-1-on-r}
\gamma_1 \sqrt{\frac{\max\{\|\bar{\nu}\|_{L_2},1\} \  r \log(en/r)}{n}} \leq \theta \sqrt{\rho},
\end{equation}
and let $\bar{k}$ to be the largest integer that satisfies
\begin{equation} \label{eq:cond-1-on-bar-k}
\gamma_1 \sqrt{\frac{\max\{\|\bar{\nu}\|_{L_2},1\} \ k \log(en/k)}{n}} \leq 2 \theta \sqrt{\rho}.
\end{equation}
If
$$
\lambda \geq 4|\E \nu|+c_1 \gamma_1\sqrt{\log(en/\bar{k})} \cdot \max\left\{\|\bar{\nu}\|_{L_2},  \frac{\|w\|_2}{\sqrt{n}}\right\},
$$
then
\begin{equation} \label{eq:Z-wv-est}
\left|\E_{\delta \otimes \nu \otimes \tau} Z_{w,v} - \frac{\inr{w,v}}{n \lambda} \right| \leq \frac{c_2}{\lambda} \left(|\E \nu|+ \theta^2 \rho \left(1+ \frac{\|w\|_2}{\sqrt{n}}\right) \right) \frac{\|v\|_2}{\sqrt{n}}.
\end{equation}
\end{Theorem}
Before we begin with the proof of Theorem \ref{thm:Z-wv-est}, let us note a few facts that follow from the growth property \eqref{eq:growth}, and in particular from \eqref{eq:monotone-single}:
\begin{Lemma} \label{lemma:growth-outcome}
There is an absolute constant $c$ for which the following holds. If $x \in \R^n$ satisfies \eqref{eq:growth} and $r \leq \ell \leq k$,
$$
\sum_{i=\ell}^k x_i^* \leq c\gamma_1 \frac{k\sqrt{\log(en/k)}}{\sqrt{n}} \ \ \ {\rm and} \ \ \ \Bigl(\sum_{i=\ell}^k (x_i^*)^2 \Bigr)^{1/2} \leq c \gamma_1 \sqrt{\frac{k\log(en/k)}{n}}.
$$
\end{Lemma}

\begin{Remark} \label{rem:growth}
It is straightforward to verify that if $x,y \in \R^n$ satisfy \eqref{eq:growth} and $\alpha_\ell \leq 2^{-\ell}$ then for every $k \geq r$,
$$
\sum_{\ell=k+1}^n \alpha_\ell \|x\|_{[\ell]} \|y\|_{[\ell]} \leq c 2^{-k} \gamma_1^2 \|x\|_{2} \|y\|_{2} \frac{k\log(en/k)}{n},
$$
where $c$ is an absolute constant.
\end{Remark}
We omit the standard proofs of these facts.

\vskip0.4cm

\noindent{\bf Proof of Theorem \ref{thm:Z-wv-est}.} Throughout this proof, we will slightly abuse notation and denote by $\noise$ the vector $(\nu_i)_{i=1}^n$ (rather than $(\delta_i\nu_i)_{i=1}^n$). Recall that $\tau$ is distributed uniformly in $[-\lambda,\lambda]$ and thus, for any $y \in \R$,
\begin{equation}
\E_\tau \sign(y+\tau) =
\begin{cases}
y/\lambda & \mbox{if} \ |y| \leq \lambda,
\\
\IND_{\{y > \lambda\}} - \IND_{\{y<-\lambda\}} & \mbox{otherwise}.
\end{cases}
\end{equation}
Let $\bar{\nu}_i=\nu_i - \E \nu$ and set $I=\{i : |w_i+\bar{\nu}_i| > \lambda\}$ (which is a random set that depends on $w$ as well). Taking the expectation with respect to $(\delta_i)_{i=1}^n$ and using that $\delta=m/n$,
\begin{align*}
\E_{\delta \otimes \tau } Z_{w,v} & =  \frac{1}{n}\sum_{i=1}^n  v_i \E_\tau \sign (w_i+\nu_i+\tau_i)
\\
& = \frac{1}{n} \sum_{i \in I^c} v_i \frac{w_i+\nu_i}{\lambda} + \frac{1}{n} \sum_{i \in I} v_i\left(\IND_{\{w_i+\nu_i > \lambda\}}-\IND_{\{w_i+\nu_i < - \lambda\}} \right)
\\
& = \frac{1}{n} \sum_{i=1}^n v_i \frac{w_i+\nu_i}{\lambda} + \frac{1}{n} \sum_{i \in I} v_i\left(- \frac{w_i+\nu_i}{\lambda} + \IND_{\{w_i+\nu_i > \lambda\}}-\IND_{\{w_i+\nu_i < - \lambda\}} \right).
\end{align*}
Taking the expectation $\E_\nu$ (i.e., with respect to $(\nu_i)_{i=1}^n$) consider the resulting two terms. Firstly,
\begin{equation} \label{eq:simple-1}
\E_\nu \frac{1}{n} \sum_{i=1}^n v_i \frac{w_i+\nu_i}{\lambda} = \frac{\inr{w,v}}{n \lambda} + \frac{\E \nu}{n \lambda} \sum_{i=1}^n  v_i \leq  \frac{\inr{w,v}}{n \lambda} + \frac{1}{\lambda} |\E\nu| \frac{\|v\|_2}{\sqrt{n}}.
\end{equation}
Secondly, let us turn to the more difficult term,
$$
\left| \E_\nu \frac{1}{n} \sum_{i \in I} v_i \left( -\frac{w_i+\nu_i}{\lambda} + \IND_{\{w_i+\nu_i > \lambda\}}-\IND_{\{w_i+\nu_i < - \lambda\}} \right) \right| =(\triangle).
$$
Using that $\IND_{\{\alpha>\lambda\}}\leq |\alpha|/\lambda$ for any $\alpha\in \R$, it follows that for every $1 \leq i \leq n$,
$$
\left| - v_i \left( \frac{w_i+\nu_i}{\lambda} + \IND_{\{w_i+\nu_i > \lambda\}}-\IND_{\{w_i+\nu_i < - \lambda\}} \right) \right| \leq 3|v_i| \cdot  \frac{|w_i+\nu_i|}{\lambda};
$$
hence,
$$
(\triangle) \leq \frac{3}{\lambda n} \E_\nu  \sum_{i \in I} |v_i| \cdot |w_i + \nu_i| \leq \frac{3 |\E \nu|}{\lambda n} \sum_{i \in I} |v_i| +  \frac{3}{\lambda n} \E_\nu  \sum_{i \in I} |v_i| \cdot |w_i + \bar{\nu}_i|.
$$
Clearly,
$$
\frac{3 |\E \nu|}{\lambda n} \sum_{i \in I} |v_i| \leq \frac{3}{\lambda} |\E \nu| \frac{\|v\|_2}{\sqrt{n}},
$$
and all that is left is to control
$$
\frac{3}{\lambda n} \E_\nu  \sum_{i \in I} |v_i| \cdot |w_i + \bar{\nu}_i|.
$$
To that end, it is standard to verify that if $(a_i)$ and $(b_i)$ are sequences then $\bigl| \sum a_i b_i \bigr|  \leq \sum a_i^* b_i^*$. Therefore,
\begin{align*}
& \E_\nu \sum_{i \in I} |v_i| \cdot |w_i+\bar{\nu}_i| \leq \E_\nu \Bigl(\sum_{i \in I} |v_i| |w_i| + \sum_{i \in I} |v_i| |\bar{\nu}_i| \Bigr) \leq \sum_{\ell=1}^n \E_\nu \Bigl(\IND_{\{|I|=\ell\}} \sum_{i=1}^\ell v_i^* \cdot (w_i^*+\bar{\nu}_i^*)\Bigr)
\\
\leq & \sum_{\ell=1}^n  \Bigl( \sum_{i=1}^\ell (v_i^*)^2 \Bigr)^{1/2} \Bigl( \sum_{i=1}^\ell (w_i^*)^2 \Bigr)^{1/2} \bP_{\nu} (|I| = \ell) + \sum_{\ell=1}^n  \Bigl( \sum_{i=1}^\ell (v_i^*)^2 \Bigr)^{1/2} \E_\nu \Bigl[\IND_{\{|I|=\ell\}}\Bigl( \sum_{i=1}^\ell (\bar{\nu}_i^*)^2 \Bigr)^{1/2}\Bigr]
\\
= & \sum_{\ell=1}^n \|v\|_{[\ell]}\|w\|_{[\ell]}\bP_{\nu} (|I| = \ell) + \sum_{\ell=1}^n \|v\|_{[\ell]} \E_{\nu} \Bigl[\IND_{\{|I|=\ell\}}\|\bnoise\|_{[\ell]}\Bigr]
\\
=& (*) + (**).
\end{align*}
Estimating $(*)$ and $(**)$ requires some preparation. Recall that $\bar{k}$ is the largest integer for which
$$
\gamma_1 \sqrt{\frac{ \max\{\|\bar{\nu}\|_{L_2},1\} \ k \log(en/k)}{n}} \leq 2\theta \sqrt{\rho},
$$
implying in particular that $\bar{k} \geq r$; hence, by \eqref{eq:monotone-single}, for $x=w$ or $x=v$,
$$
x_{\bar{k}}^* \leq  \gamma_1 \sqrt{\log(en/\bar{k})} \frac{\|x\|_2}{\sqrt{n}}.
$$
Also,
$$
\lambda \geq 2 \gamma_1 \sqrt{\log(en/\bar{k})} \frac{\|w\|_2}{\sqrt{n}}
$$
and therefore,
$$
w_{\bar{k}}^* \leq  \frac{\lambda}{2}.
$$
Next, if $\ell \geq 2\bar{k}$ there are at most $\ell/2$ indices $i$ for which $|w_i| \geq \lambda/2$, and therefore, the event
$\{ |I| = \ell \} = \left\{ \left| \left\{ i : |w_i +\bar{\nu}_i | \geq \lambda\right\} \right| = \ell \right\}$
is contained in the event
$$
{\mathcal C}_\ell = \left\{ \left| \left\{ i : |\bar{\nu}_i| \geq \frac{\lambda}{2} \right\} \right| \geq \frac{\ell}{2} \right\}.
$$
By a standard binomial estimate, for every $\ell \geq 2\bar{k}$,
\begin{equation} \label{eq:probab-c-ell}
\bP( {\mathcal C}_\ell) \leq \binom{n}{\ell/2} \bP^\ell (|\bar{\nu}| \geq \lambda/2) \leq \exp(-c^\prime(L) \ell \log(en/\ell))
\end{equation}
provided that $\lambda \gtrsim_L \|\bar{\nu}\|_{L_2}\sqrt{\log(en/\ell)}$, which is the case, again using that $\ell \geq 2\bar{k}$.

Finally, if $\ell \leq 2\bar{k}$ then for $x=w$ or $x=v$,
$$
\|x\|_{[\ell]} \leq 2\|x\|_{[\bar{k}]} \leq 4\theta \sqrt{\rho} \|x\|_2.
$$

\vskip0.3cm

Consider the term $(**)$. Since $\bnoise=(\bar{\nu}_i)_{i=1}^n$ has iid $L$-subgaussian coordinates,
\begin{equation*}
(\E \|\bnoise\|_{[\ell]}^2)^{1/2} \leq c(L)\|\bar{\nu}\|_{L_2} \sqrt{\ell \log(en/\ell)}.
\end{equation*}
Hence,
\begin{align*}
& \sum_{\ell=1}^{2\bar{k}} \|v\|_{[\ell]} \E_{\nu} \Bigl[\IND_{\{|I|=\ell\}}\|\bnoise\|_{[\ell]}\Bigr] \leq \|v\|_{[2\bar{k}]} \sum_{i=1}^{2\bar{k}} \E_\nu \Bigl[\IND_{\{|I|=\ell\}}\|\bnoise\|_{[\ell]}\Bigr]
\\
\leq & \|v\|_{[2\bar{k}]} \E_\nu \Bigl[ \|\bnoise\|_{[2\bar{k}]} \IND_{\{|I| \leq 2\bar{k}\}}\Bigr] \leq \|v\|_{[2\bar{k}]} \E \|\bnoise\|_{[2\bar{k}]} \leq c(L) \|\bar{\nu}\|_{L_2} \|v\|_{[\bar{k}]} \sqrt{\bar{k} \log(en/\bar{k})}
\\
\leq & c(L)  {\|{\bar \nu}\|_{L_2}} \gamma_1 \bar{k} \log(en/\bar{k}) \frac{\|v\|_2}{\sqrt{n}}.
\end{align*}

Turning to the sum on $\ell \in [2\bar{k},n]$,
\begin{align*}
& \sum_{\ell=2\bar{k}+1}^{n}  \|v\|_{[\ell]} \E_\nu \Bigl[\IND_{\{|I|=\ell\}} \|\bnoise\|_{[\ell]} \Bigr] \leq \sum_{\ell=2\bar{k}+1}^{n} \|v\|_{[\ell]} (\E \|\bnoise\|_{[\ell]}^2)^{1/2} \cdot \bP_\nu^{1/2}(|I|=\ell)
\\
\leq & c(L) \sum_{\ell=2\bar{k}+1}^{n} \|v\|_{[\ell]}  \|\bar{\nu}\|_{L_2} \sqrt{\ell \log(en/\ell)} \bP_\nu^{1/2}(|I| = \ell)=(\triangle \triangle).
\end{align*}
 By \eqref{eq:probab-c-ell} it is evident that $\bP_\nu^{1/2}(|I| = \ell) \leq \exp(- c^\prime(L) \ell \log(en/\ell))$, and by Remark \ref{rem:growth}
$$
(\triangle \triangle) \leq c(L) \|\bar{\nu}\|_{L_2} \gamma_1 \frac{\|v\|_2}{\sqrt{n}}  \bar{k}\log(en/\bar{k}) \exp(-c^\prime(L) \bar{k} \log(en/\bar{k})) \leq c(L) \|\bar{\nu}\|_{L_2} \gamma_1 \frac{\|v\|_2}{\sqrt{n}};
$$
thus
\begin{equation} \label{eq:(b)}
(**) \leq c(L) \|\bar{\nu}\|_{L_2} \gamma_1 \bar{k} \log(en/\bar{k}) \frac{\|v\|_2}{\sqrt{n}} \leq c(L) \theta^2 \rho \sqrt{n} \|v\|_2.
\end{equation}

The estimate on $(*)$ follows the same path, by splitting the sum to $\ell \in [1,2\bar{k}]$ and $\ell \in [2\bar{k}+1,n]$. Indeed,
\begin{align*}
& \sum_{\ell=1}^{2\bar{k}}  \Bigl( \sum_{i=1}^\ell (v_i^*)^2 \Bigr)^{1/2} \Bigl( \sum_{i=1}^\ell (w_i^*)^2 \Bigr)^{1/2} \bP_{\nu} (|I| = \ell) = \sum_{\ell=1}^{2\bar{k}} \|v\|_{[\ell]} \|w\|_{[\ell]} \bP_\nu (|I|=\ell)
\\
\leq & \|v\|_{[2\bar{k}]} \|w\|_{[2\bar{k}]} \leq c(L) \theta^2 \rho \|v\|_2 \|w\|_2,
\end{align*}
and
\begin{align*}
& \sum_{\ell=2\bar{k}+1}^n  \|v\|_{[\ell]} \|w\|_{[\ell]} \bP_{\nu} (|I| = \ell) \lesssim_L  \sum_{\ell=2\bar{k}}^n  \|v\|_{[\ell]} \cdot \|w\|_{[\ell]} \exp(-c^\prime(L) \ell \log(en/\ell))
\\
\lesssim_L &   \|v\|_{2} \cdot \|w\|_{2} \frac{\bar{k} \log(en/\bar{k})}{n} \cdot \exp(-c^\prime(L) \bar{k} \log(en/\bar{k})),
\end{align*}
using Remark \ref{rem:growth} once again. Hence,
\begin{equation} \label{eq:(a)}
(*) \leq c(L) \|v\|_2 \|w\|_2 \theta^2 \rho.
\end{equation}
Therefore, combining \eqref{eq:simple-1}, \eqref{eq:(b)} and \eqref{eq:(a)} it follows that
$$
\left|\E_{\delta \otimes \nu \otimes \tau } Z_{w,v} - \frac{\inr{w,v}}{n \lambda} \right| \leq \frac{c(L)}{\lambda} \left(|\E \nu| + \theta^2 \rho \left(1+ \frac{\|w\|_2}{\sqrt{n}}\right) \right) \frac{\|v\|_2}{\sqrt{n}},
$$
as claimed.
\endproof

\subsection{Proof of \eqref{eq:est-on-(1)-2}}
In what follows set $1 \leq r \leq n$; let $W \subset \R^n$ be a set that satisfies
$$
\log|W| \leq \gamma_2 r \log(en/r)
$$
for a suitable constant $\gamma_2$; and assume that every $w \in W$ satisfies the growth property \eqref{eq:growth} with constants $r$ and $\gamma_1$. The goal is to obtain an estimate that holds uniformly for every $w \in W$ on
\begin{equation} \label{eq:est-max-beta}
\frac{1}{m} \sup_{u \in \eta B_2^n} \max_{|I| \leq \beta m} \sum_{i \in I} \delta_i |w_i+u_i|,
\end{equation}
where $\eta$ is very small.

The idea behind the proof is that the set $W$ is well-behaved: on the one hand, its cardinality is reasonable, and on the other hand, the growth property \eqref{eq:growth} implies that vectors in $W$ are `well-spread', making them friendly to the application of selectors. Because we are interested in small perturbations of vectors in $W$ by vectors whose Euclidean norm is at most $\eta$, the impact of the perturbations is negligible.

\begin{Theorem} \label{thm:selectors-simple}
There exist absolute constants $c_1$ and $c_2$ such that the following holds. Let $W$ be as above, set $0<\beta<1$  and assume that
\begin{equation} \label{eq:cond-selector-1}
m \geq \frac{r \log^{3/2}(en/r)}{\beta}.
\end{equation}
Then with probability at least $1-2\exp(-c_1 \min\{\gamma_2 r \log(en/r),\beta m\})$ for every $w \in W$,
\begin{equation*}
\frac{1}{m} \sup_{u \in \eta B_2^n} \max_{|I| \leq \beta m} \sum_{i \in I} \delta_i |w_i+u_i| \leq \frac{\eta \sqrt{n}}{m} + c_2  \gamma_1 \gamma_2 \beta \sqrt{\log(e/\beta)}  \frac{\|w\|_2}{\sqrt{n}}.
\end{equation*}
\end{Theorem}

\proof Clearly, by the triangle inequality, for every $w \in W$,
\begin{equation*}
\sup_{u \in \eta B_2^n} \max_{|I| \leq \beta m} \sum_{i \in I} \delta_i |w_i+u_i| \leq \max_{|I| \leq \beta m} \sum_{i \in I} \delta_i |w_i| + \sup_{u \in \eta B_2^n} \|u\|_1
\leq  \max_{|I| \leq \beta m} \sum_{i \in I} \delta_i |w_i| + \eta \sqrt{n}.
\end{equation*}

Fix $w \in W$ and without loss of generality assume that its coordinates $w_i$ are nonnegative and non-increasing. Let $r$ be as in \eqref{eq:growth} and recall that $\beta m \geq r \log^{3/2}(en/r)$. Set $I_1=\{1,...,r\}$ and $I_2=\{\beta m,...,2\beta m /\delta\}$, and since $|I_1 \cup I_2|=2\beta m/\delta$, Chernoff's inequality implies that with probability at least $1-2\exp(-c\beta m)$,
$$
|\{i : \delta_i =1\} \cap \{1,...,2\beta m/\delta\} | \geq \beta m.
$$
On that event,
$$
\max_{|I| \leq \beta m} \sum_{i \in I} \delta_i w_i \leq \sum_{i \in I_1} w_i + \sum_{i \in I_2} \delta_i w_i.
$$
Clearly,
$$
\sum_{i \in I_1} w_i \leq  \sqrt{r} \|w\|_{[r]} \leq \gamma_1 r  \sqrt{\log(en/r)} \cdot \frac{\|w\|_2}{\sqrt{n}}.
$$
As for the second term, by the growth property \eqref{eq:growth}, for every $i \in I_2$
$$
w_i \leq w_r \leq \frac{\|w\|_{[r]}}{\sqrt{r}} \leq \gamma_1 \sqrt{\log(en/r)} \cdot \frac{\|w\|_2}{\sqrt{n}};
$$
recalling that $\beta m/\delta = \beta n$,
$$
\Bigl(\sum_{i \in I_2} w_i^2\Bigr)^{1/2} \leq \|w\|_{[\beta m/\delta]} \leq \gamma_1 \sqrt{n} \sqrt{\beta \log(e/\beta)} \cdot \frac{\|w\|_2}{\sqrt{n}}.
$$
By Bernstein's inequality, for $x>0$, with probability at least $1-2\exp(-x)$
\begin{align*}
\sum_{i \in I_2} \delta_i w_i \leq & \delta \sum_{i \in I_2} w_i + c\Bigl(\sqrt{x \delta} \Bigl(\sum_{i \in I_2} w_i^2\Bigr)^{1/2} + x \max_{i \in I_2} w_i \Bigr)
\\
\leq & c\gamma_1 \frac{\|w\|_2}{\sqrt{n}} \left(\delta n \cdot \beta \sqrt{\log(e/\beta)} + \sqrt{x} \sqrt{\delta n} \sqrt{\beta \log(e/\beta)} + x \sqrt{\log(en/r)} \right),
\end{align*}
where Lemma \ref{lemma:growth-outcome} is used to estimate $\sum_{i \in I_2} w_i$. Setting $x \sim \gamma_2 r \log(en/r) \geq 2 \log |W|$, it follows from the union bound  that with probability at least $1-2\exp(-c^\prime \gamma_2 r \log(en/r))$, for every $w \in W$,
\begin{align*}
& \frac{1}{m} \max_{|I| \leq \beta m} \sum_{i \in I} \delta_i |w_i| \leq \frac{\eta \sqrt{n}}{m}
\\
&\qquad \qquad  + c \gamma_1 \gamma_2 \frac{\|w\|_2}{\sqrt{n}} \cdot \frac{1}{m} \left( m \beta \sqrt{\log(e/\beta)} +  \sqrt{m}\sqrt{r \log (en/r)} \sqrt{\beta \log(e/\beta)} + r \log^{3/2}(en/r) \right)
\\
 & \leq \frac{\eta \sqrt{n}}{m} + c \beta \sqrt{\log(e/\beta)} \cdot \gamma_1 \gamma_2 \frac{\|w\|_2}{\sqrt{n}},
\end{align*}
where the last inequality holds because $\beta m \geq r \log^{3/2}(en/r)$.
\endproof

\vskip0.3cm

The following is an immediate outcome of Theorem \ref{thm:selectors-simple}.
\begin{Theorem} \label{thm:est-on-1-sparse}
There exist absolute constants $c_0,c_1$ and $c_2$ for which the following holds. Assume that
\begin{itemize}
\item[$(1)$] $\Gamma(\Sigma_{2s,n}) \subset W+\eta B_2^n$ where $W\subset \Gamma(\Sigma_{2s,n})$ satisfies $\log |W| \leq \gamma_2 s \log(en/s)$;
\item[$(2)$] Every $w \in W$ satisfies the growth property \eqref{eq:growth} with constants $2s$ and $\gamma_1$;
\item[$(3)$] For every $t \in \Sigma_{2s,n}$, $\|\Gamma t\|_2/\sqrt{n} \leq 2\|t\|_2$;
\item[$(4)$] $\beta \sqrt{\log(e/\beta)} \leq (c_0/\gamma_1 \gamma_2) \cdot (\rho/\lambda)$; and
\item[$(5)$] $m \geq c_1 \frac{ s\log^{3/2}(en/s)}{\beta}$.
\end{itemize}
Then with probability at least $1-2\exp(-c_2 \gamma_2 s \log(en/s))$,
$$
\frac{1}{m} \sup_{y \in \rho \Sigma_{2s,n}} \max_{|I| \leq \beta m} \sum_{i \in I} \delta_i |(\Gamma y)_i| \leq \frac{\eta \sqrt{n}}{m} + \frac{\rho^2}{32 \lambda}
$$
\end{Theorem}

\subsection{Proof of \eqref{eq:est-on-(2)-1}} \label{sec:est-2-1-sparse}

The key component in the proof of \eqref{eq:est-on-(2)-1} is as follows:
\begin{Theorem} \label{thm:selectors-main}
There exist constants $c_0,c_1,c_2,$ and $c_3$ that depend only on $L$ for which the following holds. Consider $W, V \subset \R^n$ that satisfy the growth property \eqref{eq:growth} with constants $r$ and $\gamma_1$, and are such that $\log |W|, \ \log |V| \leq \gamma_2 r \log(en/r)$.
Assume further that
$$
\sup_{v \in V} \frac{\|v\|_2}{\sqrt{n}} \leq c_0 \rho, \ \ \ \sup_{w \in W} \frac{\|w\|_2}{\sqrt{n}} \leq c_0.
$$
Let
$$
\eta_W \leq \min\left\{c_1 \frac{\rho}{\gamma_1 \gamma_2 \sqrt{\log(\lambda \gamma_1 \gamma_2 /\rho)}}, \frac{\lambda}{2} \right\},
$$
and set
$$
m \geq C(\gamma_1,\gamma_2) \frac{\lambda^2 r \log^{3/2}(en/r)}{\rho^2},
$$
where $C(\gamma_1,\gamma_2) =  c_2 \gamma_1^2 \gamma_2^2 \sqrt{\log \gamma_2}$.

Then with probability at least
$$
1-2\exp(-c_3\gamma_2 r \log(en/r))
$$
we have that
$$
\max_{w \in W} \sup_{u \in \eta_W B_2^n} \max_{v \in V} \sup_{u^\prime \in \eta_V B_2^n}  \left|\frac{1}{m}\sum_{i=1}^n \eps_i \delta_i (v_i+u_i^\prime) \cdot \sign (w_i + u_i + \nu_i + \tau_i) \right| \leq \frac{\rho^2}{32 \lambda} + \frac{\eta_V \sqrt{n}}{m}.
$$
\end{Theorem}

\vskip0.4cm
While the estimate in Theorem \ref{thm:selectors-main} looks rather unpleasant, one should keep in mind that in the case that interests us, $\gamma_1$ and $\gamma_2$ are poly-logarithmic in $r$ and $n$, and so is $\lambda$. Also, the factors $\eta_V$ and $\eta_W$ are very small, of the order of $n^{-2}$, and as a result terms involving them are negligible. With that in mind, the outcome of Theorem \ref{thm:selectors-main} is that
$$
\max_{w \in W} \sup_{u \in \eta_W B_2^n} \max_{v \in V} \sup_{u^\prime \in \eta_V B_2^n}  \left|\frac{1}{m}\sum_{i=1}^n \eps_i \delta_i (v_i+u_i^\prime) \cdot \sign (w_i + u_i + \nu_i + \tau_i) \right| \leq \frac{\rho^2}{16 \lambda}
$$
provided that
$$
m \geq \gamma \frac{r \log(en/r)}{\rho^2}
$$
where $\gamma$ is poly-logarithmic in $r$ and $n$.

\vskip0.4cm

The proof of Theorem \ref{thm:selectors-main} follows the same path as that of Theorem \ref{thm:selectors-simple}: reducing the wanted estimate to a bound on
$$
\max_{w \in W}  \max_{v \in V}  \left|\frac{1}{m}\sum_{i=1}^n \eps_i \delta_i v_i \cdot \sign (w_i + \nu_i + \tau_i) \right|
$$
which is handled by the union bound, taking into account the $\exp(2\gamma_2 r \log(en/r))$ pairs $(w,v)$. To achieve this reduction, one has to control the contribution of all possible $u \in \eta_W B_2^n$ and $u^\prime \in \eta_V B_2^n$. The nontrivial component in that task is identifying the random sets of signs
$$
\mathbb{S}_{w} = \left\{\sign (w_i + u_i + \nu_i + \tau_i)_{i \in I} : u \in \eta B_2^n \right\},
$$
where $I=\{i : \delta_i =1\}$. Because $w+u+\nu_{\rm noise}+\tau_{\rm thres}$ is a small perturbation of $w+\nu_{\rm noise}+\tau_{\rm thres}$, one may expect a `stability result': that on a high probability event, for every $w \in W$ the set $\mathbb{S}_w$ consists of small perturbations of the sign vector $(\sign (w_i + \nu_i + \tau_i))_{i \in I}$.
\begin{Lemma} \label{lemma:sign-perturbation}
There exist absolute constants $c_0$ and $c_1$ for which the following holds. Let $2\eta_W <\eps \leq \lambda$ and set
\begin{equation} \label{eq:cond-on-m-1}
m \geq c_0 \eps^{-1} \lambda \gamma_2 r \log(en/r).
\end{equation}
Then with probability at least $1-2\exp(-c_1 r \log(en/r))$ with respect to $(\delta_i)_{i=1}^n \otimes (\nu_i)_{i=1}^n \otimes (\tau_i)_{i=1}^n$, for every $w \in W$
$$
\mathbb{S}_w \subset \sign (w_i + \nu_i + \tau_i)_{i \in I}+2\mathcal{Z},
$$
where $I=\{i : \delta_i=1\}$, ${\mathcal Z} \subset \{-1,0,1\}^I$ and for every $z \in {\mathcal Z}$, $|{\rm supp}(z)| \leq 3\eps m /\lambda$.
\end{Lemma}

\proof Fix $\eps>0$ and note that if  $|w_i + \nu_i + \tau_i| \geq \eps$ and $|u_i| \leq \eps/2$ then
$$
\sign (w_i+u_i + \nu_i + \tau_i)=\sign (w_i + \nu_i + \tau_i).
$$
Thus, for a well chosen $\eps$ one has to show that with high probability, for every $w \in W$ and $u \in \eta B_2^n$ there are at least $(1-2\eps/\lambda)m$ coordinates $i$ such that
$$
\delta_i =1, \ \ \ |w_i+ \nu_i + \tau_i| \geq \eps, \ \  {\rm and} \ \ |u_i| \leq \eps/2.
$$
By the choice of $\eps$ one has that for every $1 \leq i \leq n$, $|u_i| \leq \|u\|_2 \leq \eta_W \leq \eps/2$; that takes care of the third constraint.

To establish the other two, recall that $\delta n =m$; that $I=\{i : \delta_i=1\}$; and that with probability at least $1-2\exp(-c^\prime m)$, $m/2 \leq |I| \leq 3m/2$. Conditioned on this event, set $(a_i)_{i \in I} \in \R^I$ to be any sequence and put $E_i=\{ |\tau_i -a_i| < \eps\}$. Note that the events $(E_i)_{i \in I}$ are independent and $\bP_\tau(E_i) \leq \eps/\lambda$; therefore, with $\tau$-probability at least $1-2\exp(-c|I|\eps/\lambda)\geq 1-2\exp(-c'm\eps/\lambda)$, there are at most $2(\eps/\lambda)|I|\leq 3(\eps/\lambda)m$ indices $i \in I$ for which $|\tau_i - a_i| < \eps$. Applying this observation to $a_i = -(w_i + \nu_i)$ conditionally on $(\nu_i)_{i=1}^n$, and then invoking a Fubini argument with respect to $(\nu_i)_{i=1}^n$ and $(\delta_i)_{i=1}^n$, it follows that for every $w \in W$, with probability at least $1-4\exp(-c' m \eps/\lambda)$ with respect to $(\delta_i)_{i=1}^n \otimes (\nu_i)_{i=1}^n \otimes (\tau_i)_{i=1}^n$,
\begin{equation} \label{eq:signs-in-proof-1}
\left|\left\{ i \in I : |w_i+\nu_i + \tau_i| \geq \eps\right\}\right| \geq \left(1-\frac{3\eps }{\lambda}\right)m.
\end{equation}
By the union bound, \eqref{eq:signs-in-proof-1} holds for every $w \in W$ provided that $\log |W| \leq cm\eps/\lambda$, which is the case by the choice of $m$.
\endproof

\vskip0.3cm

\noindent{\bf Proof of Theorem \ref{thm:selectors-main}.}  Fix $w \in W$, $u \in \eta_W B_2^n$, $v \in V$ and $u^\prime \in \eta_V B_2^n$, and note that
\begin{align*}
& \left|\frac{1}{m}\sum_{i=1}^n \eps_i \delta_i (v_i+u^\prime_i) \cdot \sign (w_i + u_i + \nu_i + \tau_i) \right|
\\
\leq & \left|\frac{1}{m}\sum_{i=1}^n \eps_i \delta_i v_i \cdot \sign (w_i + u_i + \nu_i + \tau_i) \right| + \frac{1}{m}\sum_{i=1}^n |u_i^\prime|.
\end{align*}
The second term is bounded by at most $\sqrt{n}\|u^\prime\|_2/m \leq \eta_V \sqrt{n}/{m}$. For the first term,  fix the sign vector $(\sign (w_i + \nu_i +\tau_i))_{i=1}^n$, let
$$
z_i=|\sign (w_i + u_i + \nu_i + \tau_i)-\sign (w_i + \nu_i +\tau_i)|,
$$
and set $J_z$ to be the support of $(z_i)_{i=1}^n$. Therefore,
\begin{align*}
& \left|\frac{1}{m}\sum_{i=1}^n \eps_i \delta_i v_i \cdot \sign (w_i + u_i + \nu_i + \tau_i) \right|
\\
\leq & \left|\frac{1}{m}\sum_{i=1}^n \eps_i \delta_i v_i \cdot \sign (w_i +\nu_i + \tau_i) \right| + 2 \left|\frac{1}{m}\sum_{j \in J_z} \delta_j |v_j| \right| = (a)_{w,v}+(b)_{w,v}.
\end{align*}
To estimate $(b)_{w,v}$, let ${\mathcal A}_1$ be the event from Lemma \ref{lemma:sign-perturbation} (with respect to $(\delta_i)_{i=1}^n \otimes (\nu_i)_{i=1}^n \otimes (\tau_i)_{i=1}^n$) for an $\eps$ to be specified in what follows.  Using the notation of the lemma, on the event ${\mathcal A}_1$, for every $w \in W$, $|J_z \cap I|=|{\rm supp}(z) \cap I| \leq 3\eps m /\lambda$. Setting $\beta = 3\eps/\lambda$, one has to estimate
$$
\frac{1}{m} \sum_{i \in J_z} \delta_i |v_i| = \frac{1}{m} \sum_{i \in J_z \cap I} \delta_i |v_i| \leq  \max_{|J| \leq \beta m} \frac{1}{m}\sum_{j \in J} \delta_j |v_j|,
$$
which is precisely the process studied in Theorem \ref{thm:selectors-simple} (for $\eta=0$). In particular, if $m\geq \eps^{-1}r\log^{3/2}(en/r)$, then there is an event ${\mathcal A}_2$  of probability at least
$$
1-2\exp(-c^\prime \min\{ \gamma_2 r \log(en/r), \eps m/\lambda\})
$$
with respect to $(\delta_i)_{i=1}^n$, such that for every $v \in V$,
\begin{equation} \label{eq:A-2-in-proof}
\max_{|J| \leq \beta m} \frac{1}{m}\sum_{j \in J} \delta_j |v_j| \leq  c \gamma_1 \gamma_2 \beta \sqrt{\log(e/\beta)}   \frac{\|v\|_2}{\sqrt{n}}
\sim \gamma_1 \gamma_2 \frac{\eps}{\lambda} \sqrt{\log(e \lambda/\eps)}  \frac{\|v\|_2}{\sqrt{n}}=(*).
\end{equation}
Set
\begin{equation} \label{eq:choice-of-eps}
\eps =c\frac{\rho}{\gamma_1 \gamma_2 \sqrt{\log(\lambda \gamma_1 \gamma_2/\rho)}}
\end{equation}
for a sufficiently small constant $c$, and note that by our assumption \eqref{eq:choice-of-eps} is a `legal choice' of $\eps$ (i.e., $2\eta_{W}\leq \eps$). Since $\sup_{v \in V} \|v\|_2/\sqrt{n} \leq c_1 \rho$, it is evident that
$$
(*) \leq \frac{\rho^2}{64 \lambda}
$$
with probability at least $1-2\exp(-c^\prime \gamma_2 r \log(en/r))$, as the choice of $m$ implies that $\gamma_2 r \log(en/r) \leq \eps m /\lambda$.

Finally, to estimate $(a)_{w,v}$ one may use the union bound. Indeed, conditioned on $(\nu_i)_{i=1}^n$ and $(\tau_i)_{i=1}^n$, each $w \in W$ is associated with a sign vector $(\zeta_i)_{i=1}^n$, defined by $\zeta_i=\sign (w_i+\nu_i +\tau_i)$. Therefore, as a random variable with respect to $(\eps_i)_{i=1}^n$ and $(\delta_i)_{i=1}^n$,
$$
\left|\frac{1}{m}\sum_{i=1}^n \eps_i \delta_i v_i \cdot \sign (w_i +\nu_i + \tau_i) \right| = \left|\frac{1}{m}\sum_{i=1}^n \eps_i \delta_i |v_i| \sign (v_i) \zeta_i \right|,
$$
and there are at most $|W| \cdot |V| \leq \exp(2\gamma_2 r \log(en/r))$ pairs $(v,\zeta)$. For each pair,  $(\eps_i \zeta_i \sign(v_i))_{i=1}^n$ has the same distribution as $(\eps_i)_{i=1}^n$. Without loss of generality one may assume that the $v_i$'s are nonnegative and non-increasing. Hence,
$$
\left|\sum_{i=1}^n \eps_i \delta_i v_i \right| \leq \sum_{i=1}^r |v_i| + \left|\sum_{i=r+1}^n \delta_i \eps_i v_i\right| \leq \sqrt{r} \|v\|_{[r]} + \left|\sum_{i=r+1}^n \delta_i \eps_i v_i\right|.
$$
For $i \geq r$, $v_i \leq \|v\|_{[r]}/\sqrt{r}$, so by Bernstein's inequality, with probability at least $1-\exp(-x)$,
$$
\left|\sum_{i=r+1}^n \delta_i \eps_i v_i\right| \lesssim \sqrt{\delta x}\|v\|_2 + x\frac{\|v\|_{[r]}}{\sqrt{r}}.
$$
Setting $x \sim \gamma_2 r \log(en/r)$ and invoking the union bound, it follows that with probability at least $1-2\exp(-c^\prime \gamma_2 r \log(en/r))$ with respect to $(\delta_i)_{i=1}^n \otimes (\eps_i)_{i=1}^n$, every pair $(v,w)$ satisfies
\begin{align*}
& \left|\frac{1}{m}\sum_{i=1}^n \eps_i \delta_i v_i \cdot \sign (w_i +\nu_i + \tau_i) \right|
\\
& \qquad \leq c \gamma_1 \gamma_2  \left(\frac{r \sqrt{\log(en/r)}}{m} +  \sqrt{\frac{r \log(en/r)}{m}} + \frac{r \log^{3/2} (en/r)}{m} \right) \cdot \frac{\|v\|_2}{\sqrt{n}},
\end{align*}
where we have used the growth property \eqref{eq:growth} to estimate $\|v\|_{[r]}$.

By the choice of $m$ and since $\sup_{v \in V} \|v\|_2 \lesssim \rho \sqrt{n}$, a Fubini argument shows that there is an event ${\mathcal A}_3$ with probability at least $1-2\exp(-c^\prime \gamma_2 r \log(en/r))$, such that for every $v \in V$ and $w \in W$,
$$
\left|\frac{1}{m}\sum_{i=1}^n \eps_i \delta_i v_i \cdot \sign (w_i +\nu_i + \tau_i) \right| \leq \frac{\rho^2}{64 \lambda}.
$$
The claimed estimate holds on the intersection of the events ${\mathcal A}_1$, ${\mathcal A}_2$, and ${\mathcal A}_3$ and this completes the proof.
\endproof

\section{Properties of $\Gamma_\xi$} \label{sec:circulnat}

In the previous section we have accumulated various conditions on the matrix $\Gamma$ that ensure that regardless of the identity of the sparse target $x$, any solution $x^{\#}$ of \eqref{eqn:progIsomorphicIntro} satisfies that $\|x-x^{\#}\|_2 \leq \rho$. The proofs show that to recover any $s$-sparse vector it suffices that the matrix $\Gamma$ satisfies the following properties for $r=2s$:
\begin{itemize}
\item[$(M1)$] \emph{Decomposition:} $\Gamma(\Sigma_{r,n}) \subset W + \eta B_2^n$, where $W\subset \Gamma(\Sigma_{r,n})$; $\log |W| \leq \gamma_2 r \log(en/r)$; each vector in $W$ satisfies the growth property with constants $r$ and $\gamma_1$; and $\eta$ is very small, say $\eta \lesssim 1/n^2$.

\vskip0.3cm
\item[$(M2)$] \emph{Small-ball property:} that for every $t \in \Sigma_{r,n}$, $\|\Gamma t\|_2 /\sqrt{n} \geq \kappa  \|t\|_2$.
\vskip0.3cm

\item[$(M3)$] \emph{Isomorphic upper estimate:} that for every $t \in \Sigma_{r,n}$, $\|\Gamma t\|_2/\sqrt{n} \leq \kappa^\prime  \|t\|_2$.
\end{itemize}

\begin{Remark}
Note that the combination of $(M2)$ and $(M3)$ implies that $\Gamma/\sqrt{n}$ acts on $\Sigma_{r,n}$ in an isomorphic way. It does not imply an almost isometric estimate since the constants $\kappa$ and $\kappa^\prime$ need not be close to one.
\end{Remark}

\vskip0.4cm

To complete the proof of Theorem~\ref{thm:isomorphic}, let us show that all the necessary estimates are true with high probability for a circulant matrix generated by an isotropic $L$-subgaussian random vector that has iid coordinates. The proofs of $(M2)$ and $(M3)$ follow directly from the methods developed in \cite{MRW16}. $(M1)$ can also be derived using \cite{MRW16}, though the proof  presented in what follows is somewhat simpler than the analogous claim from \cite{MRW16}.

\subsection{ $(M2)$ and $(M3)$}

To establish the small-ball property and the isomorphic upper estimate we require three facts. Let $j_0$ satisfy that $2^{j_0} = \theta (n/r)$ where $0<\theta<1$ is a suitable (small) absolute constant, and $j_1$ satisfies that $2^{j_1} = \gamma_2 r \log(en/r)$ for
$$
\gamma_2 \sim \max\left\{1,\frac{\log(er)}{\log(en/r)}\right\}.
$$

Let $T = \Sigma_{r,n} \cap S^{n-1}$ and consider $T_{j_1}, T_{j_0} \subset T$ such that $\log |T_{j_0}| \leq 2^{j_0}$ and $\log |T_{j_1}| \leq 2^{j_1}$. For every $t \in T$ let $\pi_{j_1} t \in T_{j_1}$ and put $\pi_{j_0} t \in T_{j_0}$.

\begin{Theorem} \label{thm:structure-1}
Set $r \leq cn/\log^4n$ for a suitable absolute constant $c$. There are subsets $T_{j_0}, T_{j_1} \subset T$ and maps $\pi_{j_0}$ and $\pi_{j_1}$ as above for which the following holds.  With probability at least $1-2\exp(-c^\prime 2^{j_1})$, for every $t \in \Sigma_{r,n}$,
\begin{itemize}
\item $\|\Gamma_\xi (t-\pi_{j_1}t)\|_2 \leq c^{\prime \prime}/n^2$;
\end{itemize}
with probability at least $1-2\exp(-c^\prime\min\{2^{j_0},2^{j_1}\})$, for every $t \in \Sigma_{r,n} \cap S^{n-1}$,
\begin{itemize}
\item $\left| \|\Gamma_\xi \pi_{j_1}t\|_2^2 - \|\Gamma_\xi \pi_{j_0}t\|_2^2 \right| \leq c^{\prime \prime} \sqrt{n} \sqrt{r} \alpha_r \log(e r) \leq n/8$; and
\item $\left| \|\Gamma_\xi \pi_{j_0} v\|_2^2 - n \right| \leq n/16+c^{\prime \prime} \sqrt{n} \sqrt{r} \leq n/8$,
\end{itemize}
where
$$
\alpha_r =\max\left\{1, \log \left(c \frac{r}{n^2} \log r\right) \right\}.
$$
The constants $c^\prime$ and $c^{\prime \prime}$ depend only on $L$.
\end{Theorem}

\begin{Remark}
In what follows we assume that $j_1 \geq j_0$. When $j_1 \leq j_0$ the proofs are much simpler and one may set $T_{j_0}=T_{j_1}$.
\end{Remark}

Clearly, Theorem \ref{thm:structure-1} implies the wanted two-sided isomorphic estimate. Firstly, by homogeneity, it suffices to prove the estimate in $\Sigma_{r,n} \cap S^{n-1}$. Secondly, it is standard to verify that with probability at least $1-2\exp(-cn)$, $\sup_{v \in S^{n-1}} \|\Gamma_\xi v\|_2 \leq n$. Therefore, with probability at least $1-2\exp(-c^\prime \min\{2^{j_0},2^{j_1}\}))$, for every $t \in \Sigma_{r,n} \cap S^{n-1}$,
$$
\|\Gamma_\xi t\|_2^2 \geq \|\Gamma_\xi \pi_{j_1}t \|_2^2 - 2(\sup_{v \in S^{n-1}} \|\Gamma_\xi v\|_2) \|\Gamma_\xi (t - \pi_{j_1}t)\|_2 \geq \|\Gamma_\xi \pi_{j_1}t \|_2^2 - \frac{2}{n}
$$
and
$$
\|\Gamma_\xi \pi_{j_1}t \|_2^2 \geq \|\Gamma_\xi \pi_{j_0}t\|_2^2 - \frac{n}{8} \geq \frac{3n}{4},
$$
implying that
\begin{equation} \label{eq:lower-isomorphic-in-proof}
\frac{\|\Gamma_\xi t\|_2^2}{n} \geq \frac{1}{2}  = \frac{\|t\|_2^2}{2}.
\end{equation}
The reverse direction follows in an identical manner.

\vskip0.4cm

Most of the proof of Theorem \ref{thm:structure-1} can be found in \cite{MRW16}. The proof of the first part of Theorem \ref{thm:structure-1} is a minor modification of Lemma 4.4 in \cite{MRW16}: the set $T_{j_1}$ is a net in $\Sigma_{r,n} \cap S^{n-1}$ with respect to the $\ell_2$ norm, and its cardinality---$\exp(\gamma_2 r \log(en/r))$ for $\gamma_2$ that is logarithmic in $n$ and $r$---suffices to ensure that the mesh-width of the net is $\sim 1/n^2$; in fact, the mesh-width can be improved to any power $n^{-\zeta}$ by multiplying $\gamma_2$ by a suitable constant. The proof of the second part of Theorem \ref{thm:structure-1} follows from a chaining argument with respect to a certain $\ell_\infty$-type norm---see Theorem 4.7 and Corollary 4.10 in \cite{MRW16}. The third part of Theorem \ref{thm:structure-1} is based on the following concentration result, which is a straightforward consequence of a subgaussian version of the Hanson-Wright inequality (see, for example, \cite[Lemma 5.1]{DJR17}): that for any $t\in S^{n-1}$ with $\|t\|_1\leq \sqrt{r}$ and $u>0$,
$$
\bP\left( \left| \|\Gamma_\xi t\|_2^2 - n \right| \geq u \right) \leq 2\exp\left(-c^\prime\min\left\{\frac{u^2}{rn}, \frac{u}{r} \right\}\right).
$$
Now the third part of Theorem \ref{thm:structure-1} is evident by applying this to any $t \in T_{j_0}$ with $u=n/8$ and invoking the union bound.
\endproof

\subsection{Proof of $(M1)$}

Let us show that for any $x \in \Sigma_{r,n}$, $\Gamma_\xi x$ satisfies the wanted growth property.

\begin{Theorem} \label{thm:structure-2}
For every $L, \zeta \geq 1$ there is a constant $c=c(L,\zeta)$ such that the following holds. With probability at least $1-(r/n)^{\zeta}$, for every $r \leq k \leq n$ and every $x \in \Sigma_{r,n}$,
\begin{equation}
\label{eqn:structure-2}
\|\Gamma_\xi x\|_{[k]} \leq c (\log n) \cdot (\log r) \cdot \sqrt{k \log (en/k)}\|x\|_2.
\end{equation}
\end{Theorem}

\vskip0.4cm

By combining Theorem \ref{thm:structure-2} and \eqref{eq:lower-isomorphic-in-proof}, it is evident that with probability at least $1-(r/n)^{\zeta}$ any $w \in \Gamma_\xi (\Sigma_{r,n})$ satisfies the growth property: for all $r \leq k \leq n$,
\begin{equation} \label{eq:decomp-est}
\|w\|_{[k]} \leq \gamma_1 \sqrt{k \log(en/k)} \frac{\|w\|_2}{\sqrt{n}},
\end{equation}
where $\gamma_1 = c (\log n) \cdot (\log r)$. By the first statement of Theorem~\ref{thm:structure-1}, property $(M1)$ is therefore satisfied with the choice $W=\Gamma_\xi (T_{j_1})$.

\vskip0.4cm

\proof By homogeneity it suffices to prove Theorem~\ref{thm:structure-2} for $T=\Sigma_{r,n} \cap S^{n-1}$. Just as in Theorem~\ref{thm:structure-1}, there is a set $T_{j_1} \subset T$ of cardinality at most $\exp(\gamma_2 r \log(en/r))$ and an event of probability at least $1-2\exp(-c^\prime \gamma_2 r \log(en/r))$ such that for every $t \in T$,
\begin{equation} \label{eq:approx-in-proof-1}
\|\Gamma_\xi (t - \pi_{j_1}t)\|_2 \leq \frac{c}{n^2},
\end{equation}
where $c$ is a constant that depends on $L$. Once \eqref{eqn:structure-2} is established for elements of $T_{j_1}$, it is evident that for every $t \in T$ and $r \leq k \leq n$,
\begin{align*}
\|\Gamma_\xi t\|_{[k]} \leq & \|\Gamma_\xi \pi_{j_1}t\|_{[k]} + \|\Gamma_\xi (t-\pi_{j_1}t)\|_2 \leq c (\log n)  (\log r) \sqrt{k \log (en/k)} \|\pi_{j_1} t\|_2
\\
 = & c (\log n)  (\log r) \sqrt{k \log (en/k)} \|t\|_2.
\end{align*}
To prove that the wanted estimate holds in $T_{j_1}$, recall that for every $v,x \in \R^n$, $\Gamma_\xi v = \Gamma_v \xi$ and that $\Gamma_{v+x} \xi = \Gamma_v \xi + \Gamma_x \xi$. Also, $\Gamma_v = \sqrt{n} UD_{Wv}O$ where $U,W,O$ are orthonormal matrices with entries that are bounded by $1/\sqrt{n}$ (in fact, if we denote by  $\mathcal{F}$ the discrete Fourier transform, then $U=\mathcal{F}^{-1}/\sqrt{n}$ and $W=O=\mathcal{F}/\sqrt{n}$) and $D_{Wv}$ is a diagonal matrix defined by $D_{ii} = \inr{W_i,v}$. Hence, for any $v, x \in \R^n$,
$$
\|\Gamma_v x\|_2 = \sqrt{n}\Bigl(\sum_{i=1}^n \inr{W_i,v}^2 \inr{O_i,x}^2 \Bigr)^{1/2} \leq \vertiii{v} \cdot \|x\|_2,
$$
where
$$
\vertiii{v} = \sqrt{n} \max_{1 \leq i \leq n} |\inr{W_i,v}|.
$$
Let $G$ be the standard Gaussian vector in $\R^n$, set $\| \cdot \|$ to be a norm on $\R^n$ and put $B^{\circ}$ to be the unit ball of the dual norm. Since $\xi$ is isotropic and $L$-subgaussian, a standard chaining argument shows that for any $p \geq 1$,
\begin{equation*}
\bigl(\E \|\Gamma_v \xi\|^p\bigr)^{1/p} \leq cL \bigl(\E \|\Gamma_v G\|+ \sqrt{p} \sup_{x \in B^{\circ}} \|\Gamma_v^*x\|_2 \bigr) \leq cL \bigl(\E\|\Gamma_v G\| +c\sqrt{p} \vertiii{v} \sup_{x \in B^\circ} \|x\|_2\bigr).
\end{equation*}

Fix $r \leq k \leq n$ and consider the norm $\| \cdot \|_{[k]}$. Clearly, the unit ball of the dual norm is the convex hull of $\Sigma_{k,n}$, implying that for every $v \in \R^n$,
\begin{equation} \label{eq:basic-chaining}
\bigl(\E \|\Gamma_v \xi\|_{[k]}^p\bigr)^{1/p} \leq cL \left(\E \|\Gamma_v G\|_{[k]} + \sqrt{p} \vertiii{v}\right).
\end{equation}
Observe that for every $v \in B_2^n$,
\begin{equation}
\label{eq:expec-k-largest}
\E \|\Gamma_v G\|_{[k]} \leq c \sqrt{k \log(en/k)}.
\end{equation}
Indeed, $\Gamma_v G = \sqrt{n}UD_{Wv}OG$ has the same distribution as $\sqrt{n}UD_{Wv}G$. For every $1 \leq \ell \leq n$ the random variable
$$
Z_\ell=\inr{\sqrt{n}UD_{Wv}G,e_\ell}=\sqrt{n} \sum_{i=1}^n g_i \inr{W_i,v} (U^*e_\ell)_i
$$
is a centred Gaussian random variable. Since $\|U^*e_\ell\|_\infty \leq 1/\sqrt{n}$, it follows that
$$
\|Z_\ell\|_{\psi_2} \simeq_L \|Z_\ell\|_{L_2} = \sqrt{n} \left(\sum_{i=1}^n \inr{W_i,v}^2 (U^*e_\ell)_i^2 \right)^{1/2} \leq \|Wv\|_2 \leq 1.
$$
Clearly,
$$
\E \|\Gamma_v G\|_{[k]} = \E \Bigl(\sum_{\ell \leq k} (Z_\ell^*)^2 \Bigr)^{1/2},
$$
and by a fact due to Klartag \cite{Kla02} (see also \cite[Lemma 3.5]{MRW16} for a proof),
$$
\Bigl(\E \sum_{\ell \leq k} (Z_\ell^*)^2\Bigr)^{1/2} \leq c \max_{1 \leq \ell \leq n} \|Z_\ell\|_{\psi_2} \cdot \sqrt{k\log(en/k)},
$$
implying that \eqref{eq:expec-k-largest} holds.

To complete the proof, by a well-known estimate due to Carl \cite{Car85} (see also Corollary 4.10 in \cite{MRW16}), there is a sequence of subsets $(T_j)_{j=0}^{j_1} \subset T_{j_1}$ whose cardinalities are $|T_j| \leq 2^{2^j}$,
and maps $\pi_j : T_{j_1} \to T_j$ such that for every $t \in T_{j_1}$ and every $j \leq j_1$,
$$
\vertiii{\pi_{j}t - \pi_{j-1} t} \leq c 2^{-j/2} \sqrt{r} \log(en/2^j).
$$
Set $\Delta_j t = \pi_{j+1}t - \pi_{j} t$, let $2^{\ell} = \frac{k}{r} \log(en/k)$ and assume first that $\ell \leq j_1$. Thus, for every $t \in T_{j_1}$
$$
t = \pi_{\ell} t+\sum_{j=\ell}^{j_1-1} \Delta_j t
$$
and
\begin{equation} \label{eq:chaining-1}
\sup_{t \in T_{j_1}} \|\Gamma_\xi t\|_{[k]} \leq \sup_{t \in T_{j_1}} \left(\sum_{j=\ell}^{j_1-1} \|\Gamma_{\Delta_j t} \xi \|_{[k]} + \|\Gamma_{\pi_\ell t} \xi\|_{[k]} \right).
\end{equation}
Fix $\ell \leq j \leq j_1-1$ and consider the collection of the (at most) $2^{2^{j+2}}$ random variables $\{\|\Gamma_{\Delta_j t} \xi \|_{[k]} : t \in T_{j_1}\}$. By \eqref{eq:basic-chaining} and \eqref{eq:expec-k-largest},
$$
\bigl(\E\|\Gamma_{\Delta_j t} \xi \|_{[k]}^p \bigr)^{1/p} \leq cL (\sqrt{k \log(en/k)} + \sqrt{p} \vertiii{\Delta_j t}) \leq cL \Bigl(\sqrt{k \log(en/k)} + \sqrt{\frac{p}{2^j}} \sqrt{r} \log(en/2^j) \Bigr).
$$
Setting $p \sim \zeta 2^j$, it follows from Markov's inequality and the union bound that, with probability at least $1-2\exp(-c^\prime \zeta 2^j)$, for every $t \in T_{j_1}$,
$$
\|\Gamma_{\Delta_j t} \xi \|_{[k]} \leq cL (\sqrt{k \log(en/k)} + \sqrt{\zeta} \sqrt{r} \log(en/2^j)).
$$
By the union bound the same assertion holds simultaneously for all $\ell \leq j \leq j_1$ with probability at least
$$
1-2\sum_{j=\ell}^{j_1-1} \exp(-c \zeta 2^j) \geq 1-2\exp(-c^\prime \zeta 2^{\ell})=1-2\exp(-c^{\prime \prime} \zeta (k/r)\log(en/k)).
$$
Turning to the second term in \eqref{eq:chaining-1}, observe that for $t \in \Sigma_{r,n}$, $\vertiii{t} \leq \sqrt{r}$. Therefore, by \eqref{eq:basic-chaining}, \eqref{eq:expec-k-largest}, Markov's inequality, and the union bound for the collection $\{ \Gamma_{\pi_\ell t} \xi : t \in T_{j_1} \}$, it is evident that with probability at least $1-2\exp(-c^\prime \zeta (k/r)\log(en/k))$,
$$
\|\Gamma_{\pi_\ell t} \xi \|_{[k]} \leq cL \sqrt{k \log(en/k)}\bigl(1 + \sqrt{\zeta} \log(en)\bigr).
$$
Intersecting the two events and applying the union bound for $r \leq k \leq n$, one has that with probability at least
$$
1-2\sum_{k=r}^n \exp(-c^\prime \zeta (k/r)\log(en/k)) \geq 1-\left(\frac{r}{n}\right)^{c^{\prime \prime} \zeta},
$$
for every $r \leq k \leq n$,
$$
\sup_{t \in T_{j_1}} \|\Gamma_\xi t\|_{[k]} \leq cL (j_1 - \ell) \sqrt{k \log(en/k)}\bigl(1 + \sqrt{\zeta} \sqrt{r} \log(en) \bigr),
$$
and the claim follows because $j_1 - \ell \lesssim \log r$.

The proof when $j_1 \leq \ell$ is much simpler and follows immediately from the union bound used for every $t \in T_{j_1}$ and using that $\vertiii{v} \leq \sqrt{r}$.

\endproof

\section{Proof of minimax optimality}
\label{sec:lower}

Theorem \ref{thm:lower} is established using some modifications to a more general result from \cite{Men17}.

\vskip0.4cm

\noindent{\bf Proof of Theorem \ref{thm:lower}.}
Fix $r>0$ and $0<\alpha<1$ such that $\alpha r \geq 4\rho$. Let $T \subset \Sigma_{s,n} \cap r B_2^n$ be $\alpha r$-separated with respect to the $\ell_2$-norm. Denote the rows of $A$ by $X_1,...,X_m$ (which need not be independent) and let $\mu$ be the probability distribution of $(X_i)_{i=1}^m$. Let ${\mathcal U}\subset (\R^{n})^m$ be the event
$$
\left\{ \|At\|_2 \leq \kappa \sqrt{m} r \ \ \ {\rm for \ every \ } t \in T \right\}
$$
and observe that by our assumptions, $\mu({\mathcal U}) \geq 0.95$. Denote by $\nu$ the probability distribution of $(\nu_i)_{i=1}^m$ and note that the joint distribution of $\bigl((X_i)_{i=1}^m,(\nu_i)_{i=1}^m\bigr)$ is the product measure $\mu\otimes \nu$.

Fix $x\in \Sigma_{s,n}$. Since $\Psi$ is a successful recovery procedure it follows that if $\Psi$ receives $(X_i)_{i=1}^m$ and $(\inr{X_i,x}+\nu_i)_{i=1}^m$, it outputs a vector that achieves recovery accuracy $\rho$ with confidence $0.9$; in other words,
$$
\mu \otimes \nu \left(\left\{ \bigl((X_i)_{i=1}^m,(\nu_i)_{i=1}^m\bigr) \ : \ \Psi\left( \left((X_i, \inr{X_i,x}+\nu_i)\right)_{i=1}^m \right) \in x + \rho B_2^n \right\} \right) \geq 0.9.
$$
For every $\mathbb{X} = (X_i)_{i=1}^m$ and $t_j \in T$ set
$$
\mathbb{A}_j(\mathbb{X}) := \left\{ \left(\nu_i\right)_{i=1}^m : \Psi\left( \left(X_i, \inr{X_i,t_j}+\nu_i\right)_{i=1}^m \right) \in t_j + \rho B_2^n \right\} \subset \R^m.
$$
By Fubini's Theorem and since $\mu \otimes \nu$ is a product measure, there is an event $\Omega_j\subset (\R^{n})^m$ of $\mu$-probability at least $0.8$ on which $\nu(\mathbb{A}_j(\mathbb{X})) \geq 3/4$.

Let $u_j(\mathbb{X}) = (\inr{{X}_i,t_j})_{i=1}^m$, which is simply the `noise-free' part of the measurement of $t_j$ generated by the sample ${\mathbb{X}}$. The crucial fact is that for any  ${\mathbb{X}} \in \Omega_j \cap \Omega_\ell$, the sets $u_j(\mathbb{X}) + \mathbb{A}_j({\mathbb{X}})$ and $u_\ell(\mathbb{X}) + \mathbb{A}_\ell({\mathbb{X}})$ are disjoint. Indeed, if
$$
z \in \left(u_j(\mathbb{X}) + \mathbb{A}_j({\mathbb{X}})\right) \cap \left(u_\ell(\mathbb{X}) + \mathbb{A}_\ell({\mathbb{X}})\right)
$$
then $\Psi( \mathbb{X}, z) \in t_j + \rho B_2^n$ and at the same time, $\Psi( \mathbb{X}, z) \in t_\ell + \rho B_2^n$, but those two balls do not intersect because $T$ is $4\rho$-separated.

As a result it follows that
$$
\sum_{j} \IND_{\Omega_j}(\mathbb{X}) \nu (u_j(\mathbb{X}) + \mathbb{A}_j({\mathbb{X}})) \leq 1,
$$
and setting
$$
\mathbb{B}_j({\mathbb{X}}) = -\mathbb{A}_j({\mathbb{X}}) \cap \mathbb{A}_j({\mathbb{X}}) \subset \mathbb{A}_j({\mathbb{X}}),
$$
we have
$$
\sum_{j} \IND_{\Omega_j}(\mathbb{X}) \nu \left(u_j({\mathbb{X}}) + \mathbb{B}_j({\mathbb{X}})\right) \leq 1.
$$
Integrating with respect to $\mu$,
$$
(*)=\sum_{j} \int \IND_{\Omega_j}(\mathbb{X}) \nu \left(u_j({\mathbb{X}}) +\mathbb{B}_j({\mathbb{X}})\right) \ d\mu \leq 1
$$
and all that remains is to estimate $(*)$ from below.

Recall that $\nu$ is the distribution of a Gaussian vector with mean zero and covariance $\sigma^2 I_m$. It is standard to verify (see, e.g. \cite[p.\ 82]{LeT91}) that if $K$ is a centrally symmetric subset of $\R^m$ and $y \in \R^m$ then
$$
\nu(y + K) \geq \exp(-\|y\|_2^2/2\sigma^2) \cdot \nu(K).
$$
In our case, for $\mathbb{X} \in \Omega_j$ each set $\mathbb{B}_j({\mathbb{X}})$ is centrally symmetric. Moreover, by the symmetry of $\nu$, $\nu(-\mathbb{A}_j({\mathbb{X}})) \geq 3/4$, implying that
$$
\nu(\mathbb{B}_j({\mathbb{X}})) \geq 0.5.
$$
Also, if $\mathbb{X} \in {\mathcal U}$ then $\|u_j({\mathbb{X}})\|_2 = \|A t_j\|_2 \leq \kappa \sqrt{m} r$. Note that $\mu(\Omega_j \cap {\mathcal U}) \geq 1/2$, and therefore,
\begin{align*}
(*) \geq & \frac{1}{2} \sum_{j} \int \IND_{\Omega_j}(\mathbb{X}) \exp(-\|u_j({\mathbb{X}})\|_2^2/2\sigma^2) \ d\mu \geq \frac{1}{2}\sum_{j}  \mu(\Omega_j \cap {\mathcal U}) \exp(-\kappa^2 mr^2/2\sigma^2)
\\
\geq & \frac{1}{4} |T| \exp(-\kappa^2 mr^2/2\sigma^2),
\end{align*}
It follows that if $\log |T| \geq 2\log(4)$ then
$$
m \geq \kappa^{-2} \frac{\sigma^2}{r^2} \log |T|.
$$
To complete the proof one has to show that $\Sigma_{s,n} \cap r B_2^n$ contains an $\alpha r$-separated set whose log-cardinality is at least $\sim s \log(en/s)$ for a suitable absolute constant $0<\alpha<1$, in which case one may set $r = 4\rho/\alpha$. Indeed, it is standard to verify (see, e.g., \cite[Lemma 10.12]{FoR13}) that there is a collection $\mathbb{J}$ of subsets of $\{1,...,n\}$ whose cardinality is $s$, such that $\log |\mathbb{J}| \geq cs\log(en/s)$ and $\mathbb{J}$ is $s/2$ separated with respect to the Hamming distance. For each $J \in \mathbb{J}$, let
$$
v_J = \frac{r}{\sqrt{s}} \sum_{j \in J} e_i.
$$
Then, $v_J\in \Sigma_{s,n}\cap r B_2^n$ and for $I,J \in \mathbb{J}$,
$$
\|v_I-v_J\|_2 = \frac{r}{\sqrt{s}} |I \Delta J|^{1/2} \geq \frac{r}{\sqrt{2}};
$$
thus one may set $\alpha=1/\sqrt{2}$ and the claim follows.
\endproof

\section{Extensions} \label{sec:extensions}

We conclude this article by pointing out (without providing details) some possible extensions of Theorem~\ref{thm:isomorphic} and Theorem~\ref{thm:lower}. These can be obtained by making minor modifications to the proofs presented in previous sections.

\subsection{Recovery of approximately sparse vectors}

One may extend the recovery results from sparse vectors to approximately sparse vectors. To that end, consider the recovery program
\begin{equation} \label{eqn:progIsomorphicExtend}
\max_{z\in T} \frac{1}{m}\inr{q_{\operatorname{corr}},Az} - \frac{1}{2\lambda} \frac{\|\Gamma_{\xi} z\|_2^2}{n}
\end{equation}
for the set $T=\sqrt{s} B_1^n \cap B_2^n \subset 2\co(\Sigma_{2s,n})$. Thus,
\begin{equation*}
{\rm star}(T-T) \cap \rho S^{n-1} \subset 2\sqrt{s}B_1^n \cap \rho S^{n-1}
\end{equation*}
and it is straightforward to verify that for $n\geq s/\rho^2$
\begin{equation*}
2\sqrt{s}B_1^n \cap \rho S^{n-1} \subset 4 \co(\rho \Sigma_{s/\rho^2,n}).
\end{equation*}
As a consequence, one needs to study \eqref{eq:est-on-(3)-1}, \eqref{eq:est-on-(2)-1} and \eqref{eq:est-on-(1)-2} for the pair of sets
$$
2\co(\Sigma_{2s,n}), \ \ 4 \co(\rho \Sigma_{s/\rho^2,n})
$$
rather than for the pair \eqref{eq:sets-0}. 

It is straightforward to verify that with high probability any vector $x$ in the images of the sets $\co(\Sigma_{2s,n})$ and $ \co(\Sigma_{s/\rho^2,n})$ under $\Gamma_{\xi}$ satisfies a weaker version of the growth property: that for any $r \leq k \leq n$,
\begin{equation} \label{eq:growthWeak}
\|x\|_{[k]} \leq \gamma_1  \sqrt{k\log(en/k)}
\end{equation}
where $r=2s$ or $r=s/\rho^2$, respectively, and $\gamma_1$ is a poly-logarithmic factor in $r$ and $n$. 

Using this modified growth property while following the original path of the proof, the following can be established.
\begin{Theorem} \label{thm:appSparse}
There exist constants $c_1,c_2,c_3$ depending only on $L$, and poly-logarithmic factors $\gamma_1,\gamma_2$ satisfying
$$
\gamma_1\leq \log(s/\rho^2)\log(n), \qquad \gamma_2\leq \log(n)\log\log(n)
$$
such that the following holds. Fix $0<\rho<1$; assume that $\nu$ is $L$-subgaussian and that $|\E \nu| \leq c_1 \rho$; and set $\bar{\nu}=\nu-\E\nu$. Let
$$
\lambda \geq c_2 \gamma_1 \max\{\|\bar{\nu}\|_{L_2},1\} \log(e\gamma_1^2 \max\{\|\bar{\nu}\|_{L_2},1\}/\rho)
$$
and set $\beta$ such that
$$
\beta \sqrt{\log(e/\beta)} \leq \frac{c_3}{\gamma_1 \gamma_2} \cdot \frac{\rho}{\lambda}.
$$
Let $\tau$ be uniformly distributed on $[-\lambda,\lambda]$. Put
$n\geq s/\rho^2$ and set
$$
m \geq c_3 \gamma_1^2 \gamma_2^2 \frac{\lambda^2 s \log(en/s)}{\rho^4}.
$$
Then, with probability at least $1-(\frac{s}{\rho^2 n})^{2}$, for any $x\in \R^n$ with $\|x\|_1\leq \sqrt{s}$ and $\|x\|_2\leq 1$, any solution $x^{\#}$ to \eqref{eqn:progIsomorphicIntro} satisfies $\|x^{\#}-x\|_2\leq \rho$.
\end{Theorem}
Moreover, just as in Theorem~\ref{thm:lower}, one can show that Theorem~\ref{thm:appSparse} is minimax optimal (up to a logarithmic factor).

\subsection{Heavier-tailed noise}

It is straightforward to establish a version of Theorem~\ref{thm:isomorphic} for heavier-tailed noise. In fact, it is enough that $\nu$ and $\lambda$ satisfy 
$$\bP(2|\nu|>\lambda)\leq c_1\rho; \ \ \ \E(|\nu|\IND_{\{2|\nu|>\lambda\}})\leq c_1\rho; \ \ {\rm and} \ \  |\E\nu|\leq c_1\rho.
$$

Thus, heavier-tailed noise can be compensated by stronger dithering (and, as a consequence, an increased number of measurements).

\subsection{Alternative recovery methods} \label{sec:altRec}

The outcomes of Theorem~\ref{thm:isomorphic} and Theorem ~\ref{thm:appSparse} hold for two variations of the program \eqref{eqn:progIsomorphicExtend}. Firstly, they remain valid for any solution $x^{\#}$ of
\begin{equation} \label{eqn:convTintro}
\max_{z\in T} \frac{1}{m}\langle q_{\text{corr}},Az\rangle - \frac{1}{2\la}\|z\|_2^2.
\end{equation}
Note that this program is equivalent to
\begin{equation}
\label{eqn:projTintro}
\min_{z\in T} \left\|z-\frac{\lambda}{m}A^*q_{\text{corr}}\right\|_2,
\end{equation}
i.e., its output is an $\ell_2$-projection of $\frac{\lambda}{m}A^*q_{\text{corr}}$ onto $T$. 

The program \eqref{eqn:convTintro} has some advantages: if there is a-priori knowledge that the signal $x$ is $s$-sparse and located in the Euclidean unit ball $B_2^n$, then the program can be used for $T=\Sigma_{s,n}$ and has a closed-form solution $x^{\#}$. Indeed, if $H_s$ is the hard thresholding operator then
\begin{equation}
\label{eqn:HTsol}
x^{\#}=\min\Big\{\frac{\lambda}{m},\frac{1}{\|H_s(A^*q_{\text{corr}})\|_2}\Big\} H_s(A^*q_{\text{corr}}).
\end{equation}
Secondly, the outcomes of Theorem~\ref{thm:isomorphic} and Theorem~\ref{thm:appSparse} are satisfied by any solution of the generalized Lasso
\begin{equation} \label{eqn:genLasso}
\min_{z\in T} \left\|q-\frac{1}{2\lambda}Az\right\|_2,
\end{equation}
since this program is equivalent to
$$\max_{z\in T} \frac{1}{m}\langle q,Az\rangle - \frac{1}{2\lambda} \frac{\|Az\|_2^2}{m}.$$

\subsection{Low noise regime}

In the `low noise regime', where $\|\bar{\nu}\|_{L_2}\leq c\rho$, it is possible to combine Theorem~\ref{thm:isomorphic} with adaptive thresholds during quantization (for more information on the adaptive threshold scheme, see \cite{BFN17}). This combination leads to a quantization and reconstruction scheme that, with high probability, recovers any $s$-sparse vector up to error $\rho$ from $m\geq c \gamma \log(1/\rho)s\log(en/s)$ one-bit measurements.

In a completely noiseless setting, this number of measurements is known to be optimal up to the poly-logarithmic factor $\gamma$. Since this scheme is presented in detail in the setting of subgaussian measurements in \cite[Section 3.4]{Dir18}, we will not elaborate on this further.

\end{document}